\documentclass[a4paper,11pt]{article}
\pdfoutput=1 

\usepackage{amsfonts,amssymb,amsmath}
\usepackage{jheppub} 

\usepackage[T1]{fontenc} 
\usepackage{subfigure}
\usepackage[utf8]{inputenc}

\newcommand{\x}{\mathbf{x}}

\title{Holographic quenches towards a Lifshitz point}

\author[a]{Giancarlo Camilo,}
\author[b]{Bertha Cuadros-Melgar,}
\author[a]{Elcio Abdalla}

\affiliation[a]{Instituto de F\'isica, Universidade de S\~ao Paulo\\C.P. 66318, CEP: 05315-970, S\~ao Paulo, Brasil}
\affiliation[b]{Escola de Engenharia de Lorena, Universidade de S\~ao Paulo\\Estrada Municipal do Campinho S/N, CEP: 12602-810, Lorena, Brasil}

\emailAdd{gcamilo@usp.br}
\emailAdd{bertha@usp.br}
\emailAdd{eabdalla@usp.br}

\abstract{We use the holographic duality to study quantum quenches of a strongly coupled CFT that drive the theory towards a non-relativistic fixed point with Lifshitz scaling. We consider the case of a Lifshitz dynamical exponent $z$ close to unity, where the non-relativistic field theory can be understood as a specific deformation of the corresponding CFT and, hence, the standard holographic dictionary can be applied. On the gravity side this amounts to finding a dynamical bulk solution which interpolates between AdS and Lishitz spacetimes as time evolves. We show that an asymptotically Lifshitz black hole is always formed in the final state. This indicates that it is impossible to reach the vacuum state of the Lifshitz theory from the CFT vacuum as a result of the proposed quenching mechanism. The nonequilibrium dynamics following the breaking of the relativistic scaling symmetry is also probed using both local and non-local observables. In particular, we conclude that the equilibration process happens in a top-down manner, i.e., the symmetry is broken faster for UV modes.}


\begin{document} 
\maketitle
\flushbottom

\section{Introduction}
\label{intro}

Understanding the behavior of quantum systems away from equilibrium is, in general, a challenging problem, especially when systems at strong coupling are concerned. This problem has recently attracted attention in different areas of many-body physics, motivated by recent progress in cold atom experiments that provides a way of exploring the quantum dynamics of strongly correlated systems in the laboratory \cite{2010LNP...802...21M,2011RvMP...83..863P}. A class of nonequilibrium problems of special interest in condensed matter is that of quantum quenches, i.e., the response of a quantum system to a time-dependent coupling. The typical setup consists in preparing the system at a given state (e.g., the ground state), then turning on a time-dependent coupling that approaches a constant value at late times. A key question is whether (and how) an equilibrium state is reached at the end of the process, and if such a steady state is \lq\lq thermal\rq\rq\ in any sense. Of particular interest from a theoretical point of view is the case of quenches near quantum critical points, since the response is likely to exhibit universal features that might be applied to many different physical systems. Nevertheless, the study of quench dynamics at strong coupling using standard field theory methods is usually hard and, in fact, progress in this direction has been made mostly for 2-dimensional conformal field theories (CFTs) \cite{Calabrese:2005in,Calabrese:2006rx,2010PhRvB..81a2303D,Calabrese:2007rg} (see \cite{2010AdPhy..59.1063D} for a review).

The AdS/CFT correspondence \cite{Maldacena:1997re,Gubser:1998bc,Witten:1998qj} provides a remarkable framework for the study of a certain class of strongly coupled quantum field theories by mapping them to a dual gravitational theory in higher dimensions where the treatment is classical, thus, considerably easier. It has found a wide range of applications going from quantum chromodynamics to condensed matter physics (see e.g. \cite{Adams:2012th} for a review), the majority of them in static or near equilibrium configurations. However, since the conjectured duality between the theories holds at the level of partition functions, there is no restriction on applying this framework to far from equilibrium situations as well, where it could shed light into the properties of nonequilibrium dynamics of quantum systems. On the gravity side this ammounts to studying time-dependent gravitational solutions. Indeed, holography has been used to model quenches in strongly coupled CFTs with gravity dual \cite{Albash:2010mv,Aparicio:2011zy,Basu:2011ft,Basu:2012gg,Buchel:2012gw,Liu:2013qca,Callebaut:2014tva,Rangamani:2015sha,Keranen:2011xs,Sonner:2014tca,Das:2014jna,Das:2014hqa,Buchel:2014gta,Liu:2013iza,Gao:2012aw,Garcia-Garcia:2013rha,Bai:2014tla}, where the time-dependent coupling in the CFT appears as a boundary condition for dynamical bulk fields at the boundary of anti de Sitter (AdS) space, as dictated by the holographic dictionary. Related work motivated by connections with the thermalization of the quark-gluon plasma can also be found in \cite{Danielsson:1999fa,Bhattacharyya:2009uu,Caceres:2014pda,AbajoArrastia:2010yt,Ebrahim:2010ra,Chesler:2009cy,Garfinkle:2011tc,Balasubramanian:2010ce,Balasubramanian:2011ur,Galante:2012pv,Caceres:2012em,Camilo:2014npa,Giordano:2014kya,Zeng:2013mca,Zeng:2013fsa,Zhang:2015dia,Zhang:2014cga,Fonda:2014ula,Alishahiha:2014cwa,Zeng:2014xpa,Aref'eva:2013wma,Hubeny:2013dea}. 

Whilst some effective theories in condensed matter have relativistic conformal symmetry, which is the situation where AdS/CFT is best understood, there are in turn many quantum critical points that are not conformally invariant, exhibiting instead a non-relativistic scaling (which we refer to as \emph{Lifshitz scaling}) of the form
\begin{equation}\label{eq:LifScaling}
(t,x^i)\rightarrow(\lambda^z t,\lambda x^i)\ ,
\end{equation}
where the parameter $z$ is called the dynamical critical exponent\footnote{The lack of boost invariance induced by $z$ should not sound surprising since in typical condensed matter systems there is a preferred frame set by the rest frame of the atomic lattice.}. Examples include phase transitions with $z=2$ and $z=3$ at the onset of antiferromagnetism and ferromagnetism in certain fermion systems, respectively \cite{Hartnoll:2009sz}. By following the original AdS/CFT logic of matching global symmetries of the gauge theory with isometries of the metric on the gravity side, the following Lifshitz spacetime was proposed in \cite{Kachru:2008yh,Taylor:2008tg} as a candidate background for the holographic dual of such a non-relativistic theory
\begin{equation}
 ds^2=-\frac{r^{2z}}{l^{2z}}dt^2+\frac{r^{2}}{l^{2}}d\x^2+\frac{l^{2}}{r^{2}}dr^2\ ,
\end{equation}
where the scaling symmetry (\ref{eq:LifScaling}) is realized as an isometry when combined with $r\rightarrow\lambda^{-1}r$. 
Unlike the AdS spacetime, however, this is not a vacuum solution of the Einstein equations with a negative cosmological constant -- some matter content is required to support the geometry. A number of bottom-up models have been suggested in the literature giving rise to this Lifshitz solution, such as Einstein-Proca, Einstein-Maxwell-Dilaton and Einstein-$p$-form actions \cite{Kachru:2008yh,Taylor:2008tg} (see \cite{Chemissany:2014xsa} for a general analysis of holography for bottom-up Lifshitz models, including also the hyperscaling violating solutions obtained in \cite{Gath:2012pg,Gouteraux:2012yr}), or using the nonrelativistic gravity theory of  Ho\v{r}ava-Lifshitz \cite{Griffin:2012qx}. There are even some solutions of supergravity with the specific exponent $z=2$ \cite{Donos:2010tu,Chemissany:2011mb,Christensen:2013rfa}, but at the moment no satisfactory construction of such a non-relativistic version of AdS/CFT duality is known and the problem of setting up holography for non-relativistic scenarios remains open.

An interesting step in this direction was taken in \cite{Korovin:2013bua} (see also \cite{Korovin:2013nha} for the finite temperature case, and \cite{Guica:2010sw} for related work in Schr\"odinger backgrounds), where it was shown that Lifshitz geometries with $z$ close to unity, i.e., $z=1+\epsilon^2$ with $\epsilon\ll1$, can be understood as a continuous deformation of AdS. This implies that no more than the standard AdS/CFT dictionary is required to set up holography for such a Lifshitz solution. In fact, they showed that this particular class of Lifshitz spacetime is the holographic dual of a nonrelativistic theory which is a specific deformation of the relativistic CFT corresponding to $z=1$. Namely, it is the theory obtained by deforming the CFT with the time component of a vector primary operator $\mathcal{V}^a$ of conformal dimension $\Delta=d$,
\begin{equation}\label{eq:LifAsADeformation}
 S_{\textrm{Lif}} = S_\textrm{CFT}+\sqrt{2}\epsilon\int d^dx \mathcal{V}^t(x)\ .
\end{equation}
Notice that $\epsilon$ appears here as a dimensionless small coupling constant, suggesting that conformal perturbation theory can be used to check calculations on the field theory side.
After establishing the holographic dictionary for this new class of holographic theories using an Einstein-Proca model in the bulk, the authors have checked that Lifshitz invariance indeed holds at the quantum level (to order $\epsilon^2$) and have provided a general field theoretical argument for the construction of such Lifshitz invariant models using the above recipe. 
Despite the operational convenience of working perturbatively in powers of $\epsilon$, there is also a possibility of application of these results since a number of theoretical models with $z$ close to one has appeared in condensed matter (see \cite{PhysRevB.26.154,PhysRevB.50.7526,1998PhRvL..80.5409Y,Herbut:2006cs,Son:2007ja,2000PhRvB..6114723H}). 

In the present work we study holographic quenches of CFTs in the framework of \cite{Korovin:2013bua} described above, as an attempt to model the dynamics following the breaking of the relativistic scaling symmetry of a CFT towards a nonrelativistic Lifshitz scaling of the type (\ref{eq:LifScaling}). The operator to be quenched according to a prescribed time-dependent profile is the vector operator $\mathcal{V}^t$ mentioned above. On the gravity side the problem translates into finding a dynamical solution to the Einstein-Proca model that flows between asymptotically AdS and Lifshitz spacetimes as time evolves. Dynamics in the bulk appears as a result of exciting in a time-dependent way a non-normalizable mode for the massive gauge field, as required (according to the standard AdS/CFT dictionary) in order to simulate the quench in the boundary CFT. 

The paper is structured as follows. In Section \ref{StaticLifshitz} we discuss the Einstein-Proca model in the bulk and review the static Lifshitz solutions of \cite{Taylor:2008tg} as well as the argument of \cite{Korovin:2013bua} for Lifshitz with $z=1+\epsilon^2$ ($\epsilon\ll1$) being a deformation of AdS. In Section \ref{Quenches} we find the dynamical bulk solution describing holographic quenches and discuss the final state of the time evolution. Section \ref{Thermalization} contains a study of the thermalization process as probed by both the evolution of horizons and the entanglement entropy, while final remarks appear in Section \ref{Conclusions}.

\section{Static Lifshitz solutions from a massive vector model}\label{StaticLifshitz}

Static solutions with Lifshitz isometries can be constructed from several gravity models. Here we will focus on the simplest one, first presented in \cite{Taylor:2008tg}, involving gravity with a negative cosmological constant and a massive vector field\footnote{However, the conventions used here are slightly different from \cite{Taylor:2008tg}. Namely, we follow \cite{Korovin:2013bua} where the fields and coordinates are conveniently rescaled with respect to \cite{Taylor:2008tg} by appropriate factors of $l$ in order to set the cosmological constant term independent of $z$: $g_{\mu\nu}\rightarrow l^2g_{\mu\nu},A_\mu\rightarrow l A_\mu,x^\mu\rightarrow lx^\mu$, with the overall factor of $l^{2d}$ absorbed into $G_{d+1}$. Then, by setting e.g. the AdS radius to unity this means that all dimensionful quantities are measured in units of $l_\textrm{AdS}.$},
\begin{equation}
\label{eq:LifshitzMassiveVectorAction}
 S=\frac{1}{16\pi G_{d+1}}\int d^{d+1}x\ \sqrt{-g}\left[R+d(d-1)-\frac{1}{4}F^{\mu\nu}F_{\mu\nu}-\frac{1}{2}M^2A_\mu A^\mu\right]\ .
\end{equation}
The Einstein and Proca equations of motion are, respectively,
\begin{subequations}\label{eq:LifEOM} 
\begin{align}
\label{eq:LifEOM1} 
R_{\mu\nu}&=-d\ g_{\mu\nu}+\frac{M^2}{2}A_\mu A_\nu+\frac{1}{2}F_{\mu}^{\ \sigma}F_{\nu\sigma}+\frac{1}{4(1-d)}F^{\rho\sigma}F_{\rho\sigma}g_{\mu\nu}\\
\label{eq:LifEOM2} 
\nabla_\mu F^{\mu\nu}&=M^2 A^\nu\ .
\end{align}
\end{subequations}
If we define 
\begin{equation}
 M^2=\frac{zd(d-1)^2}{z^2+z(d-2)+(d-1)^2}\qquad\textrm{and}\qquad l^2=\frac{z(d-1)}{M^2}=\frac{z^2+z(d-2)+(d-1)^2}{d(d-1)}\ ,
\end{equation}
the action (\ref{eq:LifshitzMassiveVectorAction}) admits a Lifshitz solution given by
\begin{subequations}\label{eq:StaticLifSolution} 
 \begin{align}
  ds^2&=-\frac{r^{2z}}{l^{2z}}dt^2+\frac{r^{2}}{l^{2}}d\x^2+\frac{l^{2}}{r^{2}}dr^2\\
  A&=\sqrt{\frac{2(z-1)}{z}}\frac{r^{z}}{l^{z}}dt\ .
 \end{align}
\end{subequations}
The Lifshitz scaling is realized for arbitrary dynamical exponent $z$ by the transformation $(t,x,r)\rightarrow(\lambda^z t,\lambda\x,\lambda^{-1} r)$. Clearly, when $z=1$ this becomes the usual relativistic scaling transformation, the gauge field vanishes, and the solution above reduces to the well known AdS$_{d+1}$ solution with unit curvature radius, $l_{AdS}\equiv l(z=1)=1$.

By the standard AdS/CFT dictionary, the presence of the massive vector field $A_\mu$ (viewed as a perturbation at the AdS critical point) in the bulk implies that the CFT dual to the action (\ref{eq:LifshitzMassiveVectorAction}) contains in its spectrum a vector primary operator $\mathcal{V}^a$ of dimension $\Delta$ given by 
\begin{equation}\label{eq:LifDeltaArbitraryz}
 \Delta=\frac{1}{2}\big[d+\sqrt{(d-2)^2+4M^2}\big]=\frac{d}{2}+\sqrt{\frac{(d-2)^2}{4}+\frac{zd(d-1)^2}{z^2+z(d-2)+(d-1)^2}}\ .
\end{equation}
The asymptotic expansion of the bulk vector field near $r=\infty$ is given in general by 
\begin{equation}\label{eq:Amuasymptoticexpansion}
 A_t(t,x^i,r)=r^{\Delta-d+1}A_t^{(0)}(t,x^i)+\cdots+r^{-(\Delta-1)}A_t^{(d)}(t,x^i)+\cdots\ ,
\end{equation}
where the non-normalizable mode $A_t^{(0)}$ is interpreted as the source for the dual operator and $A_t^{(d)}$ is related to its expectation value.
The theory also admits a Lifshitz critical point with $z>1$ provided $M^2$ takes values in the range \cite{Korovin:2013bua}
\begin{equation}\label{eq:LifMassRange}
 \frac{(d-1)^2(8-3d+4\sqrt{3d^2-6d+4})}{13d-16}<M^2\le\frac{d(d-1)^2}{3d-4}\ .
\end{equation}

We will be interested in the case where the dynamical exponent $z$ is very close to one, $z=1+\epsilon^2$, with $\epsilon\ll1$. In this case the static solution (\ref{eq:StaticLifSolution}) reads
\begin{subequations}\label{eq:StaticLifSolutionEpsilon} 
\begin{align}
 \!\!\!\!ds^2&=-r^{2}\!\left[1+2\epsilon^2\ln r+\frac{\epsilon^2}{1-d}\right]\!dt^2+r^{2}\!\left[1+\frac{\epsilon^2}{1-d}\right]\!d\x^2+\!\left[1-\frac{\epsilon^2}{1-d}\right]\!\frac{dr^2}{r^2}+\mathcal{O}(\epsilon^4)\\
 A&=\sqrt{2}\epsilon r\ dt+\mathcal{O}(\epsilon^3)\ ,
\end{align}
\end{subequations}
with the corresponding mass being
\begin{equation}
 M^2=d-1+(d-2)\epsilon^2+\mathcal{O}(\epsilon^4)\ .
\end{equation}
This means that the dual operator $\mathcal{V}^t$ has dimension
\begin{equation}
 \Delta = d+\frac{d-2}{d}\epsilon^2+\mathcal{O}(\epsilon^4)\ .
\end{equation}
The asymptotic expansion (\ref{eq:Amuasymptoticexpansion}) in this case reduces to 
\begin{equation}
 A_t(t,x^i,r)=r\left(1+\mathcal{O}(\epsilon^2)\right)A_t^{(0)}(t,x^i)+\cdots+r^{-(d-1)}\left(1+\mathcal{O}(\epsilon^2)\right)A_t^{(d)}(t,x^i)+\cdots\ ,
\end{equation}
which perfectly matches the static solution (\ref{eq:StaticLifSolutionEpsilon}) if we identify $A_t^{(0)}\equiv\sqrt{2}\epsilon+\mathcal{O}(\epsilon^3)$ and $A_t^{(d)}\equiv\mathcal{O}(\epsilon^3)$. In other words, the full static solution (\ref{eq:StaticLifSolutionEpsilon}) matches precisely the right asymptotic solution required for the standard AdS/CFT interpretation of the bulk model as a deformed CFT\footnote{It is important to notice that this does not happen for arbitrary $z$, since the asymptotic behavior $\sim r^{\Delta-d+1}$ of (\ref{eq:Amuasymptoticexpansion}) (with $\Delta$ given in (\ref{eq:LifDeltaArbitraryz})) is completely different from the exact solution $\sim r^z$ shown in (\ref{eq:StaticLifSolution}), unless $z=1+\epsilon^2$. This means that setting up holography for the Lifshitz solution with arbitrary $z$ (if possible) will require more than just the standard AdS/CFT dictionary, which we shall	 not pursue here.}. Therefore, to order $\epsilon^2$ the Lifshitz solution with $z=1+\epsilon^2$ has the holographic interpretation as a deformation of the corresponding CFT by a vector operator $\mathcal{V}^t$ of dimension $\Delta=d$ as anticipated in (\ref{eq:LifAsADeformation}), namely 
\begin{equation}\label{eq:LifAsADeformation2}
 S_{\textrm{Lif}} = S_\textrm{CFT}+\sqrt{2}\epsilon\int d^dx \mathcal{V}^t(x)\ .
\end{equation}

Before moving on to the study of holographic quenches in the next section we shall make some brief comments on the massive vector model (\ref{eq:LifshitzMassiveVectorAction}) used to construct the Lifshitz spacetime. This is a bottom-up model that captures the desired Lifshitz scaling provided the mass of the vector field is in the range (\ref{eq:LifMassRange}), but at the moment it is still unclear if a precise embedding in string theory exists. There are consistent Kaluza-Klein truncations of type IIB \cite{Maldacena:2008wh} and few other supergravities \cite{Deger:1998nm,Biran:1983iy,Kim:1985ez} that lead to massive vectors with $M^2$ in the required range, each corresponding to a specific value of $z$. Nevertheless, they all contain additional scalar fields coupled to the massive vector that cannot be set to zero and, hence, do not correspond to our model. Therefore, a top-down construction of holographic duality involving theories with Lifshitz symmetry remains obscure. As mentioned above, the standard AdS/CFT dictionary does not directly apply to such models since the geometry is not asymptotically AdS, and in fact not even the field theory dual (if any) to the Lifshitz geometry is known for arbitrary $z$.

\section{Holographic quenches and the breaking of relativistic scaling}\label{Quenches}

Motivated by the discussion of the previous section, we now study a simple dynamical mechanism for the breaking of the relavistic scaling of a CFT towards a non-relativistic Lifshitz scaling with $z=1+\epsilon^2$ ($\epsilon\ll1$). Namely, we study a quantum quench of the vector operator $\mathcal{V}^t$ in  (\ref{eq:LifAsADeformation2}) according to some prescribed quench profile $j(t)$, i.e., we consider the action (\ref{eq:LifAsADeformation2}) with a time depending coupling\footnote{For simplicity we normalize our quench profile with the factor of $\sqrt{2}\epsilon$, in such a way that when $J(v)\rightarrow1$ we get the Lifshitz solution with $z=1+\epsilon^2$, equation (\ref{eq:StaticLifSolutionEpsilon}).} $j(t)\equiv \sqrt{2}\epsilon J(t)$ which smoothly interpolates between the values $0$ (corresponding to a strongly coupled CFT) and $\sqrt{2}\epsilon$ (corresponding to the Lifshitz theory discussed above) as time evolves from $-\infty$ to $+\infty$,
\begin{equation}\label{eq:LifAsADeformationQuench}
 S = S_\textrm{CFT}+\sqrt{2}\epsilon\int d^dx\ J(t)\mathcal{V}^t(t,\mathbf{x})\ .
\end{equation}
This may provide new insights into the nonequilibrium process of reaching a Lifshitz critical point, e.g., in condensed matter systems. 

From the point of view of the dual gravitational description, all one needs to do is to consider the Einstein-Proca model (\ref{eq:LifshitzMassiveVectorAction}) and solve the equations of motion in a time-dependent setting subject to quench-like boundary conditions at $r\rightarrow\infty$. Namely, the non-normalizable mode of the bulk vector field $A_t$ must coincide with the desired quench profile $\sqrt{2}\epsilon J(t)$ (see details below). Notice that by turning on a non-normalizable mode proportional to $\epsilon$ the full bulk vector field will also be proportional to $\epsilon$, and therefore working perturbatively in $\epsilon$ (which is the only situation where a holographic interpretation of the final state Lifshitz theory is clear) is equivalent to solving the Einstein-Proca equations (\ref{eq:LifEOM}) perturbatively in powers of $A_\mu$. This is similar to the weak field collapse models studied in \cite{Bhattacharyya:2009uu,Caceres:2014pda}.

We begin by specifying our ansatz for the metric and vector field, which we do for arbitrary exponent $z$ before particularizing to the case of interest. As typical in dynamical problems (see e.g. \cite{Chesler:2009cy}), it will be useful to work with the ingoing Eddington-Finkelstein (EF) coordinate system $(v,r,\x)$, where $v$ is related to the usual $t$ coordinate appearing in (\ref{eq:StaticLifSolution}) via $dv=dt+\frac{l^{z+1}}{r^{z+1}}dr$. Notice that at the asymptotic boundary $r=\infty$ both $v$ and $t$ coincide, thus, any function $J(v)$ appearing in the bulk solution is understood as $J(t)$ for an observer living on this boundary (in particular, this will be the case for our quench profile on the CFT side). The ansatz for the metric and the vector field is 
\begin{subequations}\label{eq:LifAnsatz}
\begin{align}
 ds^2&=2h(v,r)dvdr-f(v,r)dv^2 +r^2d\x^2\\
 A(v,r)&=a(v,r)dv+b(v,r)dr\ . 
\end{align}
\end{subequations}
It involves $4$ unknown functions $f,h,a,b$ of both $(v,r)$, and clearly reduces to the static Lifshitz solution (\ref{eq:StaticLifSolution}) written in EF coordinates if the functions assume the static forms 
$$f_\textrm{Lif}(r)=\frac{r^{2z}}{l^{2z}}\ ,\quad h_\textrm{Lif}(r)=\frac{r^{z-1}}{l^{z-1}}\ ,\quad a_\textrm{Lif}(r)=\sqrt{\frac{2(z-1)}{z}}\frac{r^{z}}{l^{z}}\ ,\quad b_\textrm{Lif}(r)=-\frac{l^{z+1}}{r^{z+1}}a_\textrm{Lif}(r)\ ,$$
and of course the particular case of pure AdS follows by taking $z=1$.

The particularization to our case of interest ($z=1+\epsilon^2+\cdots$) is done by formally expanding each function in the ansatz (\ref{eq:LifAnsatz}) as a power series in $\epsilon$, i.e.,
\begin{subequations}\label{eq:LifPowerSeries}
 \begin{align}
   f(v,r)&=\sum_{n=0}^\infty f^{(n)}(v,r)\epsilon^n\\
   h(v,r)&=\sum_{n=0}^\infty h^{(n)}(v,r)\epsilon^n\\
   a(v,r)&=\sum_{n=0}^\infty a^{(n)}(v,r)\epsilon^n\\
   b(v,r)&=\sum_{n=0}^\infty b^{(n)}(v,r)\epsilon^n\ ,
 \end{align}
\end{subequations}
and then solving the equations of motion order by order in an $\epsilon$ expansion. We shall carry this expansion to leading non-trivial order for each function, which happens to be order $\epsilon^2$ as we will see, but the extension to arbitrary order can be done in a similar way. 

To solve the equations at a given order in $\epsilon$ one uses a power series ansatz in $r$ with log terms and $v$-dependent coefficients of the form
\begin{subequations}\label{eq:LifPowerSeries2}
 \begin{align}
 f^{(n)}(v,r) &= r^2\sum_{l=0}\frac{f^{(n)}_l(v)+\tilde{f}^{{(n)}}_l(v)\ln r}{r^l}\\
 h^{(n)}(v,r) &= \sum_{l=0}\frac{h^{(n)}_l(v)+\tilde{h}^{{(n)}}_l(v)\ln r}{r^l}\\ 
 a^{(n)}(v,r) &= r\sum_{l=0}\frac{a^{(n)}_l(v)+\tilde{a}^{{(n)}}_l(v)\ln r}{r^l}\\
 b^{(n)}(v,r) &= \frac{1}{r}\sum_{l=0}\frac{b^{(n)}_l(v)+\tilde{b}^{{(n)}}_l(v)\ln r}{r^l}\ .
 \end{align}
\end{subequations}
The equations of motion then become simple algebraic equations relating all the $v$-dependent coefficients above, except for the coefficients $a^{(n)}_0(v)$ and $f^{(n)}_0(v)$, which are left free. The former is an external input (responsible for simulating the quench in the dual boundary theory) and will be fixed by the boundary conditions (\ref{eq:LifBoundaryConditions}), while the latter represents a residual gauge freedom which we choose to fix such that the static result (\ref{eq:StaticLifSolutionEpsilon}) is recovered when the coupling does not vary in time (i.e., there is no quench at all)\footnote{Namely, to second order in $\epsilon$, $a^{(1)}_0(v)=\sqrt{2}J(v)$ and $f^{(2)}_0(v)=-J(v)^2/4$ (see next section).}.

Besides the ansatz, in order to solve the equations of motion one still needs to specify two more sets of data, the boundary conditions and the initial conditions. We first discuss the latter. Our initial configuration on the field theory side corresponds simply to a strongly coupled CFT in the vacuum state\footnote{Actually at this point all one can say is that the initial state corresponds to a zero temperature state of the strongly coupled CFT. The fact that such a zero temperature state is truly the vacuum is discussed in Section \ref{ssec:LifCorrelators}, where the expectation values of field theory operators are calculated.} (no deformation at all). In the bulk description this is represented by a pure AdS geometry and no gauge field, i.e.,
\begin{subequations}\label{eq:LifInitialConditions}
 \begin{align}
    f(v\rightarrow-\infty,r)&=r^2\\
    h(v\rightarrow-\infty,r)&=1\\
    a(v\rightarrow-\infty,r)&=0\\
    b(v\rightarrow-\infty,r)&=0\ .
 \end{align}
\end{subequations}
In particular, this set of conditions completely determines the zeroth order coefficients in the expansions (\ref{eq:LifPowerSeries}) to be
\begin{equation}\label{eq:LifInitialConditionsOrder0}
 f^{(0)}(v,r)=r^2,\qquad h^{(0)}(v,r)=1,\qquad a^{(0)}(v,r)=0,\qquad b^{(0)}(v,r)=0\ ,
\end{equation}
and demands that all the remaining $f^{(n)},h^{(n)},a^{(n)},b^{(n)}$ ($n\ne0$) vanish for $v\rightarrow-\infty$.

Now we turn to the boundary conditions at $r\rightarrow\infty$. For the vector field, in order to simulate a quench in the boundary field theory with quench profile $j(t)=\sqrt{2}\epsilon J(t)$, according to the AdS/CFT dictionary we must turn on the non-normalizable mode for its time component $a(v,r)$ with exactly the same profile $j(v)=\sqrt{2}\epsilon J(v)$ (remember that the time coordinates $v$ and $t$ coincide at $r=\infty$). For the metric components we impose that the geometry is asymptotically Lifshitz\footnote{Actually there is an abuse of terminology here. Strictly speaking, the metric is not asymptotically Lifshitz (in the usual sense) during the whole dynamical process, since the Lifshitz exponent $z$ in practice is evolving in time from $z=1$ to $z=1+\epsilon^2$, and hence the Lifshitz scaling is not realized in the intermediate steps. 
In a way, we are modelling a continuous breaking of the relativistic scaling symmetry due to the injection of energy in the form of a quench. 
What one really wants to ensure with such a boundary condition is that asymptotic Lifshitz behavior in its strict sense is reached in the final state at $v\rightarrow+\infty$, when the quench profile has stabilized to a constant value ($J(v)\rightarrow 1$) and the metric exhibits the usual Lifshitz isometry as in equation (\ref{eq:StaticLifSolutionEpsilon}).}. Thus, to order $\epsilon^2$, the boundary conditions read
\begin{subequations}\label{eq:LifBoundaryConditions}
 \begin{align}
  f(v,r\rightarrow\infty)&=r^2\left(1+2\epsilon^2J(v)^2\ln r+\cdots\right)\label{eq:LifBoundaryConditions1}\\
  h(v,r\rightarrow\infty)&=1+\epsilon^2J(v)^2\ln r+\cdots\label{eq:LifBoundaryConditions2}\\
  a(v,r\rightarrow\infty)&=\sqrt{2}\epsilon J(v)r+\cdots\label{eq:LifBoundaryConditions3}\\
  b(v,r\rightarrow\infty)&=0\ .\label{eq:LifBoundaryConditions4}
 \end{align}
\end{subequations}
At first sight the asymptotic Lifshitz behavior at the final state may sound conflicting with the pure AdS initial conditions (\ref{eq:LifInitialConditions}), but it should be kept in mind that we are dealing here with the case of $z$ very close to $1$, for which we have shown that the Lifshitz spacetime can be understood as a deformation of AdS.

It should be stressed that the function $J(v)$ is known from the beginning as an input from the CFT side (it models the precise way in which energy is injected into the system, causing a dynamical breaking of the relativistic scaling). In fact, it is the only responsible for introducing dynamics in the bulk. Our main goal is to solve the equations of motion (\ref{eq:LifEOM}) for the unknown functions in the ansatz (\ref{eq:LifAnsatz})-(\ref{eq:LifPowerSeries}) as functionals of $J(v)$.

\subsection{The solution to order $\epsilon^2$}

For simplicity we focus here on the case $d=3$, namely a quantum quench of a CFT living in $(2+1)$ dimensions, which is motivated by a variety of layered $2$-dimensional systems in condensed matter, but a similar analysis should hold for any number $d$ of dimensions with no additional complications. By carrying out the perturbative scheme introduced above and taking into account the initial conditions (\ref{eq:LifInitialConditions}) and boundary conditions (\ref{eq:LifBoundaryConditions}), the solution to order $\epsilon^2$ for the vector field and the metric reads
\begin{subequations}\label{eq:LifQuenchSolPart1}
 \begin{align}
   A(v,r)&=\epsilon\left[a^{(1)}(v,r)dv+b^{(1)}(v,r)dr\right]+\mathcal{O}(\epsilon^3)\\
   ds^2&=2\left[1+\epsilon^2h^{(2)}(v,r)\right]dvdr-\left[r^2+\epsilon^2f^{(2)}(v,r)\right]dv^2 +r^2\left(dx_1^2+dx_2^2\right)+\mathcal{O}(\epsilon^4)\ ,
 \end{align}
\end{subequations}  
with the functions $a^{(1)},b^{(1)},f^{(2)},h^{(2)}$ given in terms of the quench profile $J(v)$ as\footnote{If $J(v)\equiv1$ for all $v$ (i.e., the coupling is a constant and there is no quench at all), all derivatives of $J$ (and hence the coefficient $I(v)$) vanish and our solution reduces to the static Lifshitz solution with $z=1+\epsilon^2$, equation (\ref{eq:StaticLifSolutionEpsilon}). This solution has been explored in \cite{Korovin:2013bua}, where it was shown to be dual to the vacuum state of the Lifshitz field theory. As we shall see in Section \ref{ssec:LifCorrelators}, for quench profiles going asymptotically from $0$ to $1$ this interpretation is no longer true for the final state of the evolution (essentially due to the non-vanishing contribution of $I(v)$, causing a breaking of the Lifshitz symmetry), which will correspond to a thermal state.}
\begin{subequations}\label{eq:LifQuenchSolPart2}
 \begin{align}
   a^{(1)}(v,r)&=\sqrt{2}r\left(J(v)+\frac{\dot{J}(v)}{r}+\frac{\ddot{J}(v)}{2r^2}\right)\\
   b^{(1)}(v,r)&=-\frac{\sqrt{2}}{r}\left(J(v)+\frac{\dot{J}(v)}{2r}\right)\\
   f^{(2)}(v,r)&=2r^2\left(\ln r-\frac{1}{4}\right)J(v)^2-3rJ(v)\dot{J}(v)-\dot{J}(v)^2-\frac{I(v)}{r}\\
   h^{(2)}(v,r)&=J(v)^2\ln r-\frac{J(v) \dot{J}(v)}{r}-\frac{\dot{J}(v)^2}{8 r^2}\ .
 \end{align}
\end{subequations}
The coefficient $I(v)$ is defined as
\begin{equation}
 I(v)=\frac{1}{2}\int_{-\infty}^{v} \ddot{J}(w)^2\,dw\ .
\end{equation}
Unlike all the remaining coefficients, its value at instant $v$ depends on the whole history of the function $\ddot{J}^2$ integrated up to this time, and for that reason this coefficient will play a decisive role in determining the end state of the process, as we shall see in the sequence.

We begin the discussion by checking the trivial limit $v\rightarrow-\infty$, where the function $J$ and all its derivatives vanish due to our assumption that $J(v)$ asymptotes to zero. The coefficient $I(v)$ also trivially vanishes, and we are left with the static AdS solution with no gauge field, in agreement with our initial conditions (\ref{eq:LifInitialConditions}).

Now let us analyze the final state at $v\rightarrow+\infty$. We have assumed that the quench profile $J(v)$ asymptotes to the constant value $1$, so all coefficients involving derivatives of $J(v)$ will vanish except for $J(v)$ itself. In addition, the coefficient $I(v)$ approaches a constant positive value, namely
\begin{equation}\label{eq:LifIfdefinition}
I_f=\frac{1}{2}\int_{-\infty}^{\infty} \ddot{J}(w)^2dw>0\ .
\end{equation}
This means that the end state will correspond to an asymptotically Lifshitz black brane with $z=1+\epsilon^2$, namely
\begin{equation}\label{eq:LifFinalStateBB}
ds^2_f = 2\left(1+\epsilon^2\ln r\right)dvdr-r^2\left[1+2\epsilon^2\left(\ln r-\frac{1}{4}\right)-\epsilon^2\frac{I_f}{r^3}\right]dv^2 +r^2\left(dx_1^2+dx_2^2\right)+\mathcal{O}(\epsilon^4)\ ,
\end{equation}
supported by a finite vector field configuration $A=\sqrt{2}\epsilon r(dv-\frac{1}{r^2}dr)$. The corresponding event horizon will be located at $r=r_h$ given by the largest solution of 
\begin{equation}
 1+2\epsilon^2\left(\ln r_h-\frac{1}{4}\right)-\epsilon^2\frac{I_f}{r_h^3}=0\ .
\end{equation}
 
The fact that $I_f>0$ implies that it is impossible to reach a pure Lifshitz solution at the final state, since there will always occur a black hole formation. Exciting the non-normalizable mode of the vector field triggers a gravitational collapse in the bulk. From the boundary field theory point of view, this means that quenching the vector operator $\mathcal{V}^t$ in the CFT vacuum will always drive the system to a nonrelativistic Lifshitz theory \emph{at finite temperature}. Another way to state this is that it is impossible to reach the vacuum state of the Lifshitz theory from the vacuum of a CFT as result of a (continuous) quench of the operator $\mathcal{V}^t$.

The Ricci scalar for the solution (\ref{eq:LifQuenchSolPart1})-(\ref{eq:LifQuenchSolPart2}) is easily found by taking the trace of the Einstein equation (\ref{eq:LifEOM1}), namely
\begin{eqnarray}
 R &=& -d(d+1)+\frac{M^2}{2}A_\mu A^\mu+\frac{3-d}{4(1-d)}F_{\mu\nu}F^{\mu\nu}\nonumber\\
 &=&-12-2\epsilon ^2 \left[J(v)^2+\frac{2 J(v)\dot{J}(v)}{r}+\frac{J(v)\ddot{J}(v)+\frac{3}{4}\dot{J}(v)^2}{r^2}+\frac{\dot{J}(v)\ddot{J}(v)}{2r^3}\right]+O\left(\epsilon^4\right)  .
\end{eqnarray}
A curvature singularity appears at $r=0$ but, as discussed above (see also next section for two explicit examples), it is not naked since it is always covered by a horizon at $r_\textrm{EH}(v)\sim\left(\epsilon^{2}I(v)\right)^{1/3}$. This also sets the regime of validity for our perturbative solution, namely the range of values for the radial coordinate going from the boundary $r=\infty$ down to $r_h\sim \left(\epsilon^{2}I_f\right)^{1/3}$. Together with $\epsilon\ll1$ this ensures that none of the terms in the solution (\ref{eq:LifQuenchSolPart1})-(\ref{eq:LifQuenchSolPart2}) spoil the assumption of weak field and, hence, the solution can be trusted.

\subsection{Two quench profiles of interest}

We now discuss two particular quench profiles of interest, which will be used for a detailed study of observables in the sequence.

The first one is a function which interpolates between the values $0$ and $1$ in a time scale $\delta t$, such as (see e.g. \cite{Balasubramanian:2011ur})
\begin{equation}\label{eq:LifQuenchProfileTanh}
 J(v)=\frac{1}{2}\left(1+\tanh\frac{v}{\delta t}\right)\ .
\end{equation}
This is the case we have been anticipating from the beginning, in which our solution describes a dynamical geometry evolving from pure AdS to a Lifshitz black hole with $z=1+\epsilon^2$. For such a profile, the coefficient $I(v)$ (the only nontrivial coefficient in the solution (\ref{eq:LifQuenchSolPart2})) can be analytically found as being
\begin{equation}
 I(v)=\frac{2+5\tanh^3\left(v/\delta t\right)-3\tanh^5\left(v/\delta t\right)}{30 \delta t ^3}\ .
\end{equation}
In particular, its final value at $v\rightarrow+\infty$, which according to (\ref{eq:LifFinalStateBB}) is related to the mass parameter of the final state Lifshitz black hole, is 
\begin{equation}
 I_f=\frac{2}{15\delta t^3}\ .
\end{equation}
The fact that it goes with $\sim1/\delta t^3$ implies that one should be careful when using our perturbative solution for the case of a fast quench ($\delta t\rightarrow0$). As we have mentioned before, the perturbative solution is only valid for values of $r$ going from infinity up to $r\sim r_h$ (the event horizon of the final state black hole) given by $r_h\simeq\left(\epsilon^2I_f\right)^{1/3}\simeq0.5\epsilon^{2/3}/\delta t$. Therefore, for this choice of quench the perturbative solution can only be trusted deep inside the bulk provided $\epsilon\ll1$ but also $\epsilon^{2/3}/\delta t\ll2$ (or $r_h\ll1$).

A second quench profile of interest, which is slightly different from the transition we have been considering, is a Gaussian function that starts and ends asymptotically at 0, i.e., 
\begin{equation}\label{eq:LifQuenchProfileGauss}
 J(v)=e^{-v^2/2\delta t^2}\ .
\end{equation}
This means that the relativistic scaling of the CFT is broken by the quench at intermediate times but restored at the end state and it would be interesting to explore how this happens.
In the bulk description such a choice corresponds to a dynamical spacetime starting at pure AdS, evolving in time in a nontrivial way, and ending up by forming an asymptotically AdS black hole. One must have in mind that the expression for the final state black hole in this case will again have the form (\ref{eq:LifFinalStateBB}), but now without the two $\ln r$ terms which are exclusive of Lifshitz black holes. The coefficient $I(v)$ then is
\begin{equation}
 I(v)=\frac{1}{16 \delta t^6}\left[3 \sqrt{\pi}\delta t^3\left(1+\text{erf}(v/\delta t)\right)+2v e^{-v^2/\delta t^2} \left(\delta t^2-2v^2\right)\right]\ ,
\end{equation}
and the corresponding final value at $v\rightarrow+\infty$ reads
\begin{equation}
 I_f=\frac{3\sqrt{\pi}}{8\delta t^3}\ .
\end{equation}
This quantity will be related to the mass of the Schwarzschild-like AdS black hole formed at the end of the process. It again depends on the quenching time as $\sim1/\delta t^3$, so the same comment made for the first quench concerning the regime of validity applies here and, in particular, one must be careful when applying our solution if we are interested in the fast quench limit. Namely, the location of the horizon now will be $r_h=(\epsilon^2I_f)^{1/3}\simeq0.9\epsilon^{2/3}/\delta t$, hence the perturbative solution is only reliable deep inside the bulk provided $\epsilon\ll1$ as well as $\epsilon^{2/3}/\delta t\ll1$ (or $r_h\ll1$).

\subsection{All-order structure of the perturbative solution}

Although in the present work we are only interested in keeping terms up to $\epsilon^2$ in the perturbative expansion introduced in (\ref{eq:LifPowerSeries}), nothing prevents us from proceeding to higher orders in $\epsilon$ \footnote{Of course the boundary conditions (\ref{eq:LifBoundaryConditions}) must be appropriately modified in these cases.}. For the sake of completeness, here we analyze the all-order structure of the Einstein-Proca equations of motion (\ref{eq:LifEOM}). 

Since the vector field (which is turned on at order $\epsilon^1$ by the boundary condition (\ref{eq:LifBoundaryConditions3})) backreacts quadratically on the Einstein equation, it is straightforward to see that the metric will only receive contributions at even powers of $\epsilon$. As a consequence, the gauge field will contain only odd powers of $\epsilon$. In summary, the final form of the perturbative solution will look schematically like
\begin{subequations}
 \begin{align}
  f(v,r)&=r^2+\sum_{n=1}^{\infty}\epsilon^{2n}f^{(2n)}(v,r)\\
  h(v,r)&=1+\sum_{n=1}^{\infty}\epsilon^{2n}h^{(2n)}(v,r)\\
  a(v,r)&=\sum_{n=0}^{\infty}\epsilon^{2n+1}a^{(2n+1)}(v,r)\\
  b(v,r)&=\sum_{n=0}^{\infty}\epsilon^{2n+1}b^{(2n+1)}(v,r)\ .
 \end{align}
\end{subequations}

\section{Holographic probes of thermalization}\label{Thermalization}

In this section we use the gravity solution previously obtained to study the nonequilibrium dynamics of observables with a known holographic description. 
Since the solution (\ref{eq:LifQuenchSolPart1})-(\ref{eq:LifQuenchSolPart2}) fluctuates at intermediate times but always reaches a static thermal configuration after some time, as shown above, a clear notion of thermalization is ensured to happen in our model. Then, an interesting point would be to study the thermalization time of the field theory following the quench, and how this is affected at different scales. In Vaidya-like approaches to the problem of holographic thermalization (see e.g. \cite{Balasubramanian:2011ur}) the conclusion was that the UV (short distance) modes thermalize before IR (large distance) modes, the so called top-down thermalization. It would be useful to check if the same holds here, as well as the role played by the quenching rate $\delta t$.

\subsection{Correlation functions}\label{ssec:LifCorrelators}

We begin by studying the time evolution of two local observables, namely the vacuum expectation values $\langle T_{ab}\rangle$ and $\langle \mathcal{V}^a\rangle$ of the stress-energy tensor and of the quenching operator $\mathcal{V}^a$ in the boundary theory. The standard procedure to obtain correlation functions in AdS/CFT involves renormalizing the on-shell gravitational action using the holographic renormalization prescription \cite{deHaro:2000vlm}\footnote{The holographic renormalization of Lifshitz theories is usually done using the vielbein formalism, which is more appropriate to deal with non-relativistic theories (see, e.g., \cite{Ross:2009ar}). Here, however, since we are studying the Lifshitz theory as a deformation from the point of view of an AdS critical point, the standard metric formulation of \cite{deHaro:2000vlm} applies.}, then varying the renormalized action with respect to the corresponding sources (asymptotic boundary values of the corresponding bulk fields) to get the correlators.

The holographic calculation of correlation functions has been carried out in full detail for the Einstein-Proca model (\ref{eq:LifshitzMassiveVectorAction}) in \cite{Korovin:2013bua} (to order $\epsilon^2$). In Appendix \ref{LifHoloRenorm} we provide a summary of the relevant results, which are valid for an arbitrary solution to the bulk equations of motion, and discuss how to apply them to our quench solution. The resulting correlators are given by
\begin{subequations}\label{eq:LifCorrelatorsQuench}
 \begin{align}
 \big\langle \mathcal{V}^{t}(t)\big\rangle &= - \frac{\epsilon}{16\sqrt{2}\pi G}\dddot{J}(t) + \mathcal{O}(\epsilon^3)\\
 \big\langle T_{tt}(t)\big\rangle &= -\frac{\epsilon^2}{16\pi G}\left[-2I(t)+\dot{J}(t)\ddot{J}(t) - J(t)\dddot{J}(t)\right] + \mathcal{O}(\epsilon^4)\\
 \big\langle T_{ij}(t)\big\rangle &= -\frac{\epsilon^2}{32\pi G}\left[-2I(t)+\dot{J}(t)\ddot{J}(t)\right]\delta_{ij} + \mathcal{O}(\epsilon^4)\ ,
 \end{align}
\end{subequations}
with the remaining components vanishing up to higher order terms in the $\epsilon$ expansion, i.e., $\big\langle\mathcal{V}^{i}\big\rangle = \mathcal{O}(\epsilon^3)$ and $\big\langle T_{ti}\big\rangle = \mathcal{O}(\epsilon^4)$. 

\begin{figure}[htp]
    \centering
    \subfigure[Tanh quench profile (\ref{eq:LifQuenchProfileTanh})]{\includegraphics[width=7.0cm,height=5.0cm]{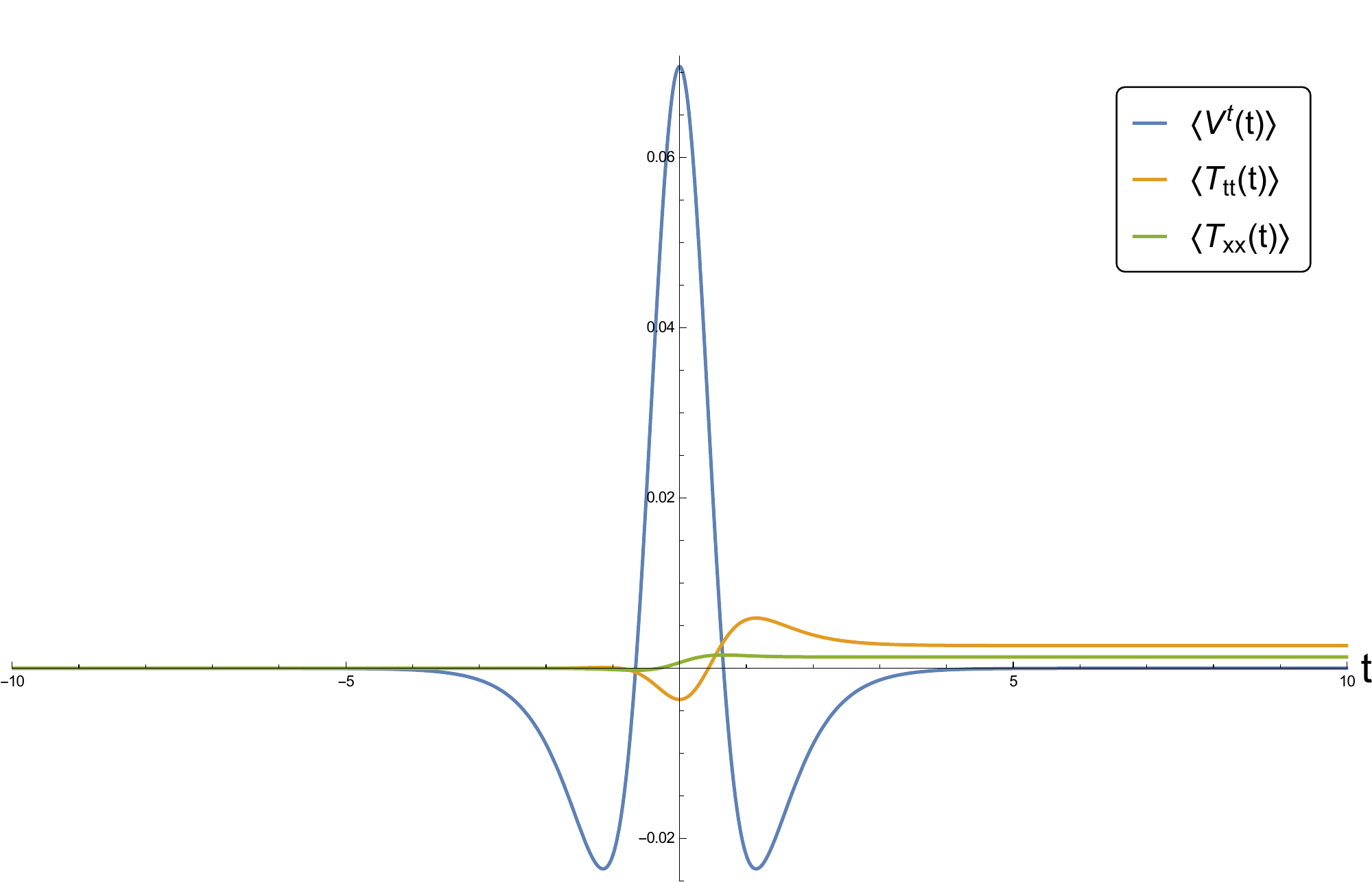}}\qquad
    \subfigure[Gaussian quench profile (\ref{eq:LifQuenchProfileGauss})]{\includegraphics[width=7.0cm,height=5.0cm]{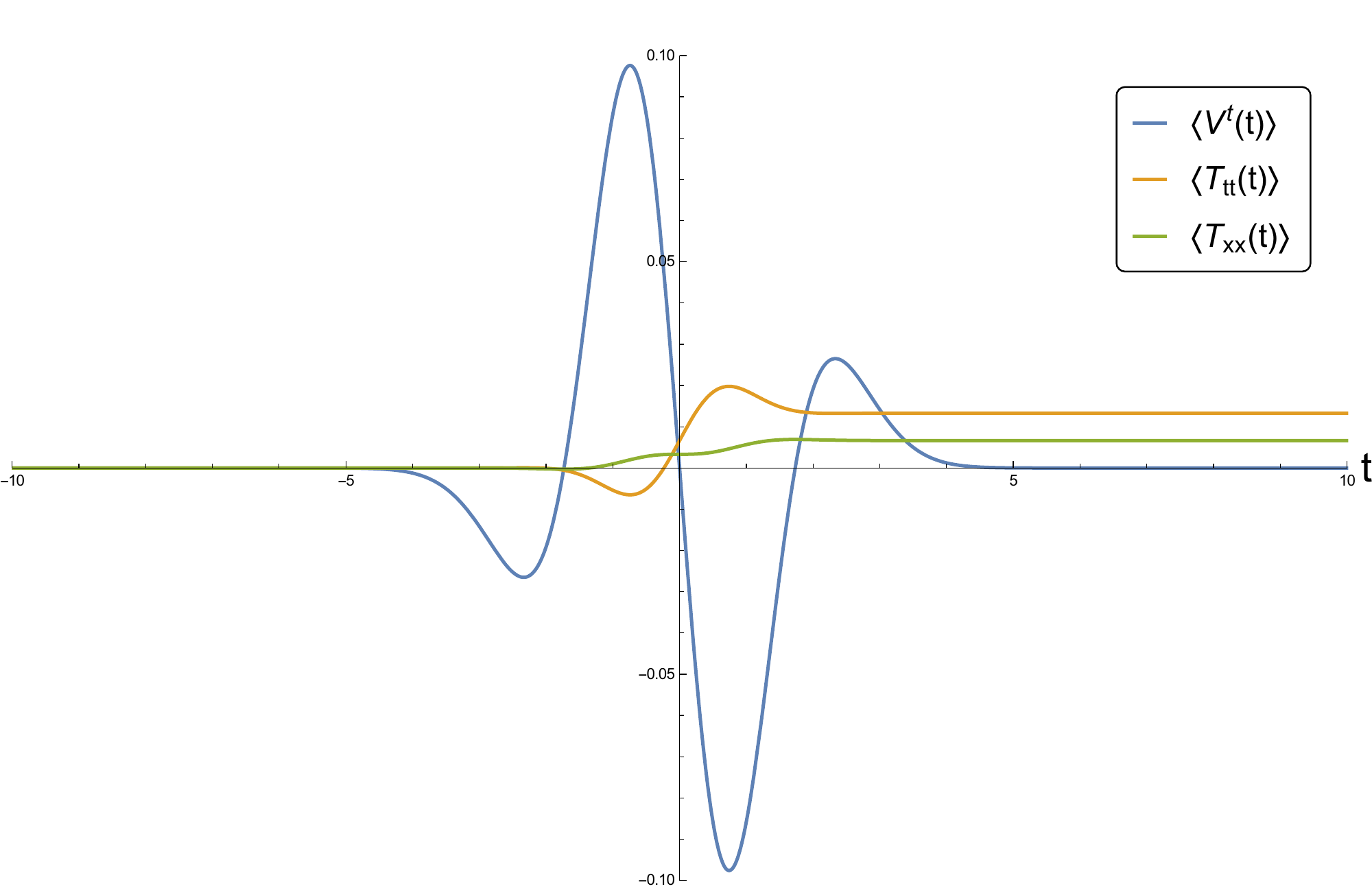}}\qquad
    \caption{\small Time evolution of the correlators (\ref{eq:LifCorrelatorsQuench}) for the two quench profiles $J(t)$ of interest. For the plots we set $16\pi G\equiv1$ and choose the values $\epsilon=0.1,\delta t=1$.}\label{fig:correlators}
\end{figure}

The time evolution of the correlators (\ref{eq:LifCorrelatorsQuench}) is shown in Figure \ref{fig:correlators} for the two quench profiles of interest. In the initial state ($t=-\infty$) they all vanish (to order $\epsilon^2$) as a consequence of our assumption that $J(t)$ asymptotes to zero at early times (see initial conditions (\ref{eq:LifInitialConditions})), where the bulk solution reduces simply to empty AdS space, i.e.,
\begin{equation}
 \big\langle \mathcal{V}^{t}(-\infty)\big\rangle = \big\langle T_{tt}(-\infty)\big\rangle = \big\langle T_{xx}(-\infty)\big\rangle = 0\ .
\end{equation}
Thus, the initial geometry can be interpreted as the vacuum state of the CFT as mentioned before. The same is not true for the final state at $t=+\infty$. Despite the fact that $J$ asymptotes to a final value and, hence, all its derivatives vanish at $+\infty$, the function $I(t)$ approaches a positive constant $I_f$ (see (\ref{eq:LifIfdefinition})). As a result, $\big\langle T_{tt}(+\infty)\big\rangle$ and $\big\langle T_{xx}(+\infty)\big\rangle$ are non-vanishing and the final state cannot correspond to the vacuum of the Lifshitz theory. Such a (\lq\lq spontaneous\rq\rq)~breaking of the Lifshitz symmetry has been anticipated before due to the appearence of a finite temperature at the end state.

The correlation functions in the boundary theory are tipically constrained by the presence of Ward identities. In fact, it follows from (\ref{eq:LifCorrelatorsQuench}) that
\begin{equation}\label{eq:LifWardQuench1}
\partial^t \big\langle T_{tt}\big\rangle = -\frac{\epsilon^2}{16\pi G}J(t)\ddddot{J}(t) + \mathcal{O}(\epsilon^4) = \sqrt{2}\epsilon J(t)\partial_t\big\langle \mathcal{V}^{t}\big\rangle + \mathcal{O}(\epsilon^4)\ .
\end{equation}
This is precisely (the time component of) the expected diffeomorphism Ward identity \cite{Bianchi:2001kw}, namely
\begin{equation}\label{eq:LifWardGeneral1}
 \nabla^b\big\langle T_{ab}\big\rangle = A_{[0]a}\nabla_b\big\langle \mathcal{V}^{b}\big\rangle - F_{[0]ab}\big\langle \mathcal{V}^{b}\big\rangle\ ,
\end{equation}
since $A_{[0]a}=\sqrt{2}\epsilon J(t)dt+\mathcal{O}(\epsilon^3)$ (see (\ref{eq:LifFeffermanGrahamAsymptoticQuench})) is the source term coming from the vector field to leading order in $\epsilon$ (the corresponding field strength $F_{[0]}=0$). If the quench profile $J(t)$ were a constant this would be just expressing the conservation of energy and momentum in the boundary theory. Therefore, in this sense, the right-hand side of (\ref{eq:LifWardQuench1}) describes the \lq\lq work\rq\rq~done on the system by varying the coupling in time.

In addition to (\ref{eq:LifWardGeneral1}) there is also the conformal Ward identity \cite{Bianchi:2001kw}
\begin{equation}
 \big\langle T^{a}_{a}\big\rangle = A_{[0]a}\big\langle \mathcal{V}^{a}\big\rangle + \mathcal{A}\ ,
\end{equation}
where $\mathcal{A}$ is the conformal anomaly. The correlators (\ref{eq:LifCorrelatorsQuench}) are trivially checked to satisfy this constraint with $\mathcal{A}=0$, which is in agreement with the well known fact that there are no conformal anomalies in odd dimensions (remember that $d=3$ in our case). However, it should be clear that tracelessness of $T_{ab}$ is only guaranteed at the initial and final states of the evolution ($\big\langle T^{a}_{a}\big\rangle\sim J\ \dddot{J}$), which is not surprising since, as mentioned before, the quench breaks the scaling symmetry at intermediate times.

\subsection{Time evolution of the apparent and event horizons}

In this section we study the time evolution of the apparent and event horizons. Although for a static black hole the two horizons necessarily coincide, this is not the case in a dynamical spacetime \cite{Booth:2005qc}. In fact, they can evolve in time in completely different ways, being coincident only when the equilibrium state is reached and the black hole is formed. 
In general, if a gravitational collapse process is sourced by a physically reasonable matter field, the apparent horizon should always lie inside the event horizon. In addition, the area of the event horizon is expected to grow monotonically during the entire process. Here we study these two features for our dynamical solution, since they provide nontrivial consistency checks of the solution.

The apparent horizon is defined as the outermost trapped surface, that is, the closed surface on which all outgoing null rays normal to it have zero expansion (i.e., they stop expanding outwards). It is a local concept in the sense that its existence can be inferred by an observer looking only at a small region of the spacetime. The notion of apparent horizon is not an invariant property of the spacetime, since its location or even its existence depends on how spacetime is foliated.
This is in sharp contrast with the concept of event horizon, defined as the null surface inside of which light rays can never escape to null infinity. Notice that the existence of an event horizon is a fundamental causal property of the spacetime which does not depend on the choice of coordinates, since determining whether or not light is able to escape to null infinity requires the knowledge of the entire history of the spacetime.

We begin by calculating the event horizon for our solution. It is defined as the null surface $\mathcal{S}(v,r)\equiv r-r_\textrm{EH}(v)=0$, meaning that its normal vector $\partial_\mu\mathcal{S}=\partial_r-r'_\textrm{EH}(v)\partial_v$ must be null, i.e., $g^{\mu\nu}\partial_\mu\mathcal{S}\partial_\nu\mathcal{S}=0$. For a spacetime of the form (\ref{eq:LifAnsatz}) this results in the following differential equation for $r_\textrm{EH}$:
\begin{equation}\label{eq:LifEventHorizon}
 \frac{dr_\textrm{EH}}{dv}=\frac{f(v,r_\textrm{EH})}{2h(v,r_\textrm{EH})}\ .
\end{equation}

In order to obtain the apparent horizon, we first need to introduce the tangent vectors $\xi^\mu_\textrm{in/out}$ to the ingoing and outgoing radial null geodesics in the spacetime (\ref{eq:LifAnsatz}). They are given by
\begin{equation}
 \xi^\mu_\textrm{in}=-\partial_r\ ,\qquad \xi^\mu_\textrm{out}=\frac{1}{h(v,r)}\left[\frac{f(v,r)}{2h(v,r)}\partial_r+\partial_v\right]\ ,
\end{equation}
where the normalization was chosen such that $\xi_\textrm{in}^2=\xi_\textrm{out}^2=0$ and $\xi_\textrm{in}\cdot\xi_\textrm{out}=-1$. Then the apparent horizon is located at the radius $r_\textrm{AH}(v)$ where the expansion $\theta_\textrm{out}(v,r)$ of a congruence of outward pointing null geodesics vanishes, namely 
\begin{equation}
 \theta_\textrm{out}=\mathcal{L}_{\xi_\textrm{out}}\ln \sqrt{-\gamma}=\xi^\mu_\textrm{out}\partial_\mu\ln \sqrt{-\gamma}\quad\equiv0\quad\textrm{for }r=r_\textrm{AH}(v)\ .
\end{equation}
Here $\mathcal{L}_{\xi_\textrm{out}}$ denotes the Lie derivative along $\xi_\textrm{out}$ (which acts just as a directional derivative for a scalar function) and $\sqrt{-\gamma}=r^2$ is the area element on the codimension-2 surface $\gamma_{ij}dx^idx^j=r^2(dx_1^2+dx_2^2)$ which is orthogonal to this null congruence. It is straightforward to show from the formulas above that $\theta_\textrm{out}=\frac{f(v,r)}{r h(v,r)^2}$, so the apparent horizon is determined by the equation
\begin{equation}\label{eq:LifApparentHorizon}
 f(v,r_\textrm{AH})=0\ .
\end{equation}

\begin{figure}[ht]
    \centering
    \subfigure[Tanh quench profile (\ref{eq:LifQuenchProfileTanh})]{\includegraphics[width=7cm,height=5.5cm]{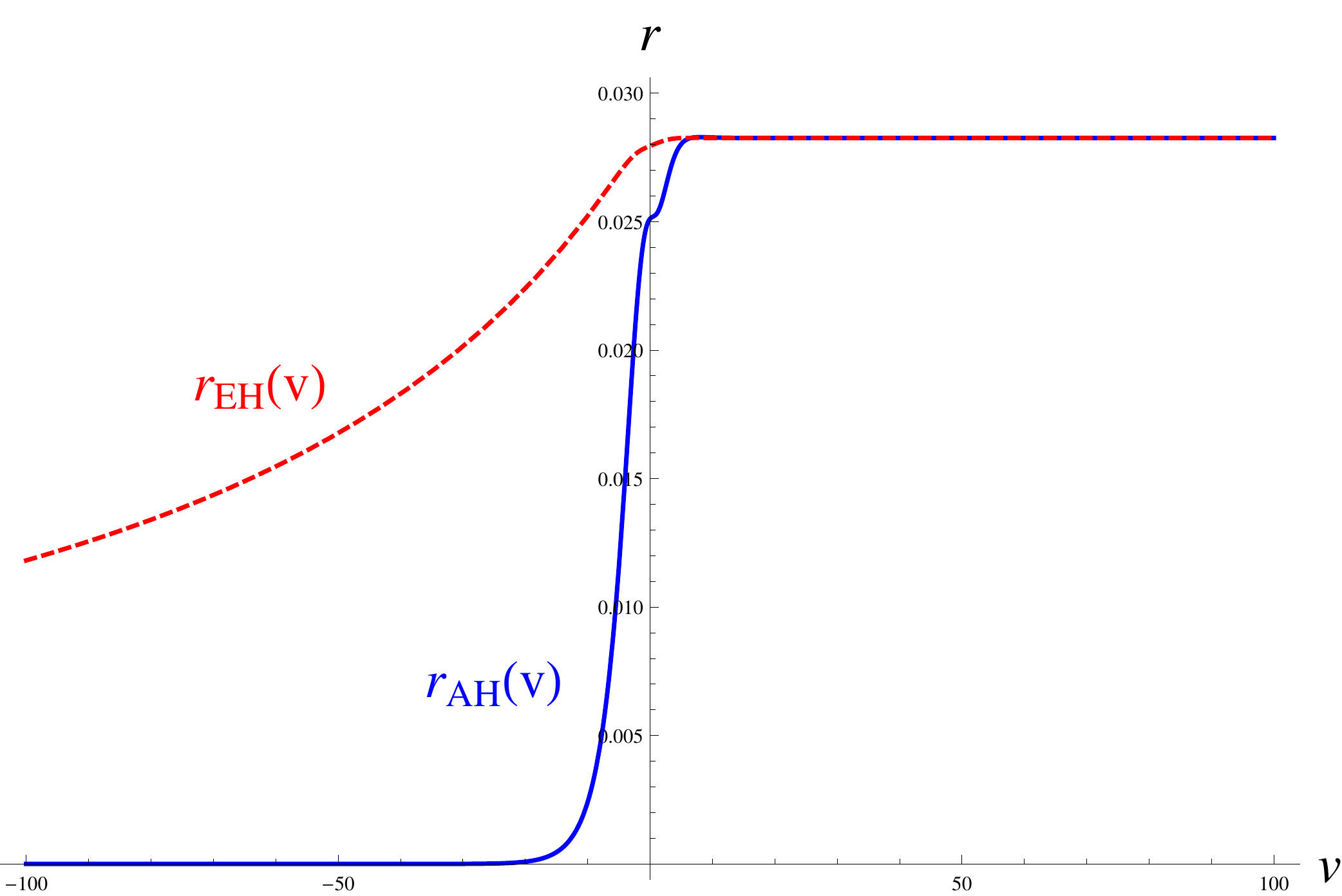}}\qquad
    \subfigure[Gaussian quench profile (\ref{eq:LifQuenchProfileGauss})]{\includegraphics[width=7cm,height=5.5cm]{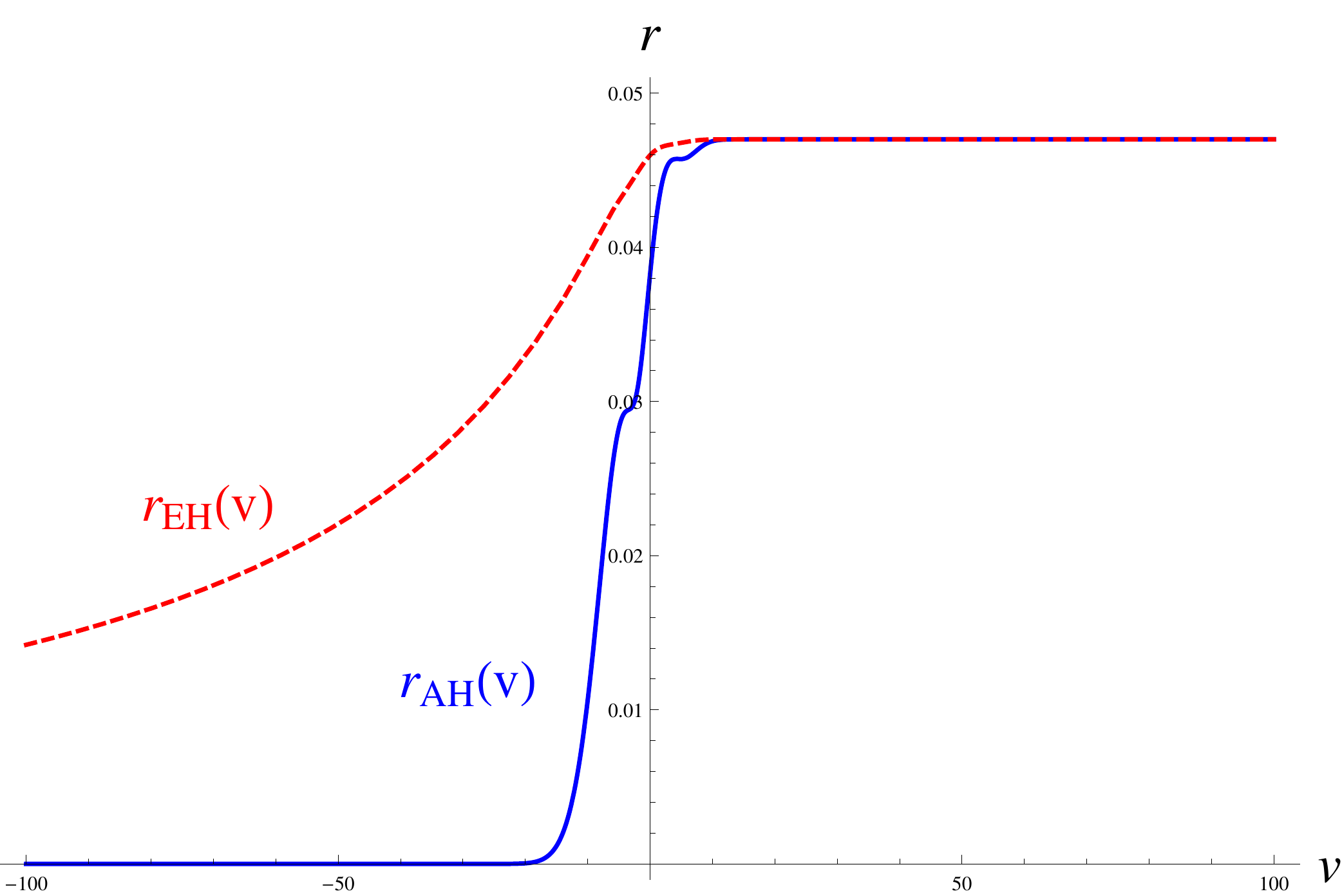}}\qquad
    \caption{\small Evolution of the apparent horizon $r_\textrm{AH}(v)$ (blue) and the event horizon $r_\textrm{EH}(v)$ (red, dashed), as dictated by equations (\ref{eq:LifApparentHorizon}) and (\ref{eq:LifEventHorizon}), respectively. For the plots we choose $\epsilon=0.1$ and $\delta t=4$. The relatively large value for $\delta t$ was chosen for didatic purposes to make evident the non-trivial behavior of $r_\textrm{AH}(v)$. As the value of $\delta t$ is decreased (faster quenches) $r_{\textrm{AH}}(v)$ approaches a step function.}\label{fig:rEHrAH}
\end{figure}

Expressions (\ref{eq:LifEventHorizon}) and (\ref{eq:LifApparentHorizon}) with $f(v,r)=r^2+\epsilon^2f^{(2)}(v,r)$ completely determine the location of the event and apparent horizons for our solution (\ref{eq:LifQuenchSolPart1})-(\ref{eq:LifQuenchSolPart2}) once the quench profile $J(v)$ is specified. In figure \ref{fig:rEHrAH} we show a comparison of $r_\textrm{EH}(v)$ and $r_\textrm{AH}(v)$ during the whole time evolution for the two quench profiles of interest. In both cases one sees that the apparent horizon lies behind the event horizon during the whole collapse process, as expected. It also follows that the area of the event horizon, which is proportional to $r_\textrm{EH}(v)^2$, will grow monotonically in time (and similarly for the area of the apparent horizon). The two horizons reach the same static values at the end of the process, as expected, and this happens at roughly the same time of order $\sim\delta t$. Therefore we see that our solution trivially passes the two consistency checks.

\subsection{Entanglement entropy}

An interesting non-local observable in field theory with a well known dual gravity description is the entanglement entropy of a spatial subregion $A$. 
For any quantum field theory in a given state $\rho$ (such as the vacuum state $|0\rangle\langle0|$) the entanglement entropy of a spacetime region $A$ with its complement $B$ provides a notion of how much entanglement exists between the two regions. It is defined as 
\begin{equation}
S_A=-\textrm{Tr}_A\left(\rho_A\ln\rho_A\right)\ , 
\end{equation}
i.e., as the von Neumann entropy associated with the reduced density matrix $\rho_A=\textrm{Tr}_B\rho$ obtained by tracing over the degrees of freedom in region $B$. 

The AdS$_{d+1}$/CFT$_d$ correspondence provides a simple and elegant way to compute the entanglement entropy in a strongly coupled gauge theory with a gravitational dual in terms of a geometrical quantity in the bulk. This so called holographic entanglement entropy formula, first proposed by Ryu and Takayanagi \cite{Ryu:2006bv} (see also \cite{Hubeny:2007xt} for the covariant version) is given by
\begin{equation}\label{eq:LifHEE}
 S_A=\frac{1}{4G_N^{(d+1)}}\textrm{ext}_{\gamma_A}(\textrm{Area}(\gamma_A))\ ,
\end{equation}
where $G_N^{(d+1)}$ is the Newton's constant in $d+1$ dimensions and $\gamma_A$ is a codimension-2 surface in the bulk with its border $\partial\gamma_A$ coinciding with the border $\partial A$ of the desired entangling region $A$ of the CFT living in the AdS boundary. The symbol $\textrm{ext}_{\gamma_A}$ denotes the extremal surface among all the $\gamma_A$'s (in the sense of \cite{Hubeny:2007xt}). In the case where the entangling region $A$ is chosen at a constant time slice (which will be our case), this condition reduces simply to finding the minimal area bulk surface with $\partial\gamma_A=\partial A$. 

As one can see from the holographic formula above, the entanglement entropy clearly depends on both the size and shape of the entangling region. This means that it can capture physical properties at many different length scales, and hence using the entanglement entropy as a probe for the quench dynamics of the CFT may be helpful to understand the equilibration process at different scales.

Now we particularize to $d=3$, which is our case of interest. For simplicity, we consider the simplest shape for the boundary entangling region $A$, namely a strip-like geometry in the $(x^1,x^2)$ directions at a constant time slice. We take the strip to have infinite width (regulated by $\ell_\perp\rightarrow\infty$) in the $x^2$ direction and a finite width $\ell$ in the $x^1$ direction. Due to this infinite extension, the entangling region is translation invariant along $x^2$ and hence the bulk surface will depend only on $x^1\equiv x$, which can be used to parametrize the functions $v(x)$ and $r(x)$ characterizing the surface.

The area functional for the class of bulk surfaces $\gamma_A$ described above becomes
\begin{equation}\label{eq:LifArea} 
A[v,r]=\ell_\perp\int_{-\ell/2}^{\ell/2}dx\ r(x)\sqrt{r(x)^2+2h(v,r)r'(x)v'(x)-f(v,r)v'(x)^2}\ ,
\end{equation}
where  $'=\frac{d}{dx}$. Notice that the infinite length $\ell_\perp$ of the $x^2$ direction factorizes and, since we are interested just in the $\ell$ dependence, we can study the density $A[v,r]/\ell_\perp$ instead of the area itself. The pair of functions $\left(v_\textrm{min}(x),r_\textrm{min}(x)\right)$ minimizing this functional will be the minimal surface $\gamma_A$ appearing in the Ryu-Takayanagi formula and, then, the holographic entanglement entropy will be $S_A=\mathcal{A}/4G_N$, where $\mathcal{A}\equiv A[v_\textrm{min},r_\textrm{min}]$.

Expanding the metric coefficients $f,h$ in powers of $\epsilon$ as dictated by the solution (\ref{eq:LifQuenchSolPart2}), and also the time and radial profiles\footnote{A comment on notation: we will omit from now on the subscript \lq\lq min\rq\rq~to denote the minimal area surface, and also denote the $\epsilon^n$ terms in the $\epsilon$ expansion of the functions $v$ and $r$ as $v_n\equiv v^{(n)},r_n\equiv r^{(n)}$ in order to keep the notation as clean as possible in the sequence.} $v(x),r(x)$ of the minimal surface as 
\begin{subequations}\label{eq:LifEEProfilesExpanded}
 \begin{align}
  v&=v_{0}+\epsilon^2v_{2}+\mathcal{O}(\epsilon^4)\\
  r&=r_{0}+\epsilon^2r_{2}+\mathcal{O}(\epsilon^4)\ , 
 \end{align}
\end{subequations}
it follows that the entanglement entropy can also be written as a power series in $\epsilon$, i.e., 
\begin{equation}
\label{eq:LifEEExpanded}
 S_A=S_A^{(0)}+\epsilon^2S_A^{(2)}+\mathcal{O}(\epsilon^4)\ .
\end{equation}
The zeroth order contribution is given by
\begin{equation}
\label{eq:LifEEOrder0}
S_A^{(0)}=\frac{\ell_\perp}{4G_N}\int_{-\ell/2}^{\ell/2}dx\ L(v_0,r_0)
\end{equation}
where we have defined (the \lq\lq Lagrangian\rq\rq\ for minimal surfaces in the background AdS spacetime)
\begin{equation}\label{eq:LifEELagrangian}
 L(v_0,r_0)=r_0\sqrt{r_{0}^2+2r_{0}'v_{0}'-r_{0}^2v_{0}'^2}\ .
\end{equation}
The second order contribution is 
\begin{eqnarray}\label{eq:LifEEOrder2}
 S_A^{(2)}&=&\frac{\ell_\perp}{4G_N}\int_{-\ell/2}^{\ell/2}dx\ \frac{r_0^2\left[2v_{0}'r_{0}' h^{(2)}(v_{0},r_{0})-v_{0}'^2 f^{(2)}(v_{0},r_{0})\right]}{2L(v_0,r_0)}\nonumber\\
 &&+\frac{\ell_\perp}{4G_N}\int_{-\ell/2}^{\ell/2}dx\ \frac{r_0^2v_{0}'r_{2}'+r_0^2(r_{0}'-r_{0}^2 v_{0}')v_{2}'+2r_0\left[r_{0}'v_{0}'+r_{0}^2(1-v_{0}'^2)\right]r_{2}}{L(v_0,r_0)}\ .
\end{eqnarray}
Notice that it depends on both the zeroth order profiles $(v_{0},r_{0})$ and second order profiles $(v_{2},r_{2})$, meaning that in order to get $S_A^{(2)}$ one needs to calculate $v_{2},r_{2}$ as well. As we shall see below, the integral in the second line contributes to the entanglement entropy only a term proportional to $r_2(\ell/2)$. Thus it is not necessary to solve for the full $r_2(x)$ (only near $x=\ell/2$, what is considerably easier).

To get $v_0$ and $r_0$ we need to solve the Euler-Lagrange equations arising from (\ref{eq:LifEEOrder0}). This can be done with the help of two immediate conserved quantities, the \lq\lq Hamiltonian\rq\rq~$H$ and the \lq\lq momentum\rq\rq~$p_{v_0}$ arising from the fact that $L(v_0,r_0)$ does not depend explicitly on $x$ and on $v_0(x)$, respectively, i.e.,
\begin{subequations}\label{eq:LifEEConserved}
 \begin{align}
 H(x)&\equiv-\frac{r_0(x)^3}{\sqrt{r_{0}(x)^2+2r_{0}'(x)v_{0}'(x)-r_{0}(x)^2v_{0}'(x)^2}}=-r_{*}^2\label{eq:LifEEConservedH}\\
 p_{v_0}(x)&\equiv\frac{r_0(x)\left[r_{0}'(x)-r_{0}(x)^2v_{0}'(x)\right]}{\sqrt{r_{0}(x)^2+2r_{0}'(x)v_{0}'(x)-r_{0}(x)^2v_{0}'(x)^2}}=0\label{eq:LifEEConservedp}\ .  
 \end{align}
\end{subequations}
Above, we have introduced the modified boundary conditions for the minimal surface at $x=0$, namely
\begin{equation}
\label{eq:LifEEModBCs}
 v_{0}(0)=v_{*},\qquad r_{0}(0)=r_{*},\qquad r_{0}'(0)=v_{0}'(0)=0\ .
\end{equation}
These follow from the fact that the surface stretching from the boundary to the bulk interior must be symmetric with respect to $x=0$, therefore this must be a turning point. Of course these are not our original boundary conditions defined by the boundary time $t$ and separation $\ell$, but it will turn out to be convenient to work with these modified boundary conditions when integrating the equations of motion. At the end we can go back and express our solution in terms of $t,\ell$ instead of $v_{*},r_{*}$ using the relations 
\begin{equation}
\label{eq:LifEEOriginalBCs}
 v_0(\pm\ell/2)=t,\qquad r_0\left(\pm\ell/2\right)=r_{\infty}\ .
\end{equation}
Here, $r_{\infty}$ is a cutoff for the AdS boundary introduced to regulate possible divergences arising due to the UV behavior of the metric.

Solving the two conservation equations (\ref{eq:LifEEConserved}) for $r_0'(x)$ and $v_0'(x)$ results in
\begin{subequations}\label{eq:LifEEDerivativesrv}
 \begin{align}
  v_0'(x)&=\frac{r_0'(x)}{r_0(x)^2}\\
  r_0'(x)&=r_0(x)^2\sqrt{\frac{r_0(x)^4}{r_*^4}-1}\ .\label{eq:LifEEDerivativesrveqb}
 \end{align}
\end{subequations}
It is not possible to integrate these equations in terms of elementary functions due to the fourth power appearing inside the square root. However, an exact solution can be obtained in terms of special functions\footnote{Although, in order to get the entanglement entropy, it is not actually necessary to integrate these equations and find the explicit form of the functions $v_0,r_0$. Namely, one could simply change the integration variable from $x$ to $r_0(x)$ inside the integral in (\ref{eq:LifEEOrder2}) using (\ref{eq:LifEEDerivativesrv}) and never worry about the exact form of $r_0(x)$ itself. Anyway, we find it instructive to present the exact form (\ref{eq:LifEEsol}).}. Taking into account the modified boundary conditions (\ref{eq:LifEEModBCs}) the solution is given as an implicit function of $x$ by
\begin{subequations}\label{eq:LifEEsol}
 \begin{align}
 v_{0}(x)&=v_*+\frac{1}{r_*}-\frac{1}{r_0(x)}\\
 x&=\frac{\sqrt{\pi}\Gamma(3/4)}{r_*\Gamma(1/4)}-\frac{r_*^2}{3 r_0(x)^3}\,
   _2F_1\left(\frac{1}{2},\frac{3}{4};\frac{7}{4};\frac{r_*^4}{r_0(x)^4}\right)
 \end{align}
\end{subequations} 
where $\Gamma(u)$ is the gamma function, $_2F_1(a,b;c;x)$ is the hypergeometric function, and the parameters $v_*,r_*$ are related to the original $t,\ell$ boundary conditions via 
\begin{equation}\label{eq:LifEEparameters}
 t=v_*+\frac{1}{r_*},\qquad\ell=\frac{2}{r_*}\frac{\sqrt{\pi}\Gamma(3/4)}{\Gamma(1/4)}=\frac{1.19814}{r_*}\ .
\end{equation}

The background contribution to the entanglement entropy, $S_A^{(0)}$, does not depend on $t$~\footnote{This follows simply from the background AdS spacetime being static, but it can also be seen explicitly from the fact the \lq\lq Lagrangian\rq\rq~(\ref{eq:LifEELagrangian}) does not depend on $v_0(x)$, which according to the solution (\ref{eq:LifEEsol})-(\ref{eq:LifEEparameters}) is the only place where $t$ appears.}, so in order to study the time evolution one subtracts this constant value and study $\delta S_A(t)=S_A(t)-S_A^{(0)}$ instead of $S_A(t)$ itself. To order $\epsilon^2$ this is given by equation (\ref{eq:LifEEOrder2}). In the integral appearing in the first line we simply change the integration variable from $x$ to $r_0(x)$ with the help of (\ref{eq:LifEEDerivativesrv}). In the second line, we first integrate the $r_{2}'$ term by parts to get a total derivative and a term proportional to $r_{2}$; then use the equations of motion (\ref{eq:LifEEDerivativesrv}) to show that the coefficients multiplying $r_{2}$ and $v_{2}'$ vanish; the only term remaining is the total derivative $\big(\sqrt{1-r_*^4/r_0^4}\ r_2\big)'$. This is trivially integrated to yield a surface term that can be simplified using the boundary condition (\ref{eq:LifEEOriginalBCs}), resulting simply in $2r_2(\ell/2)$. Therefore, the time evolution of the entanglement entropy finally becomes
\begin{eqnarray} \label{eq:LifEEFinal}
\delta S_A(t)&=&\epsilon^2\frac{\ell_\perp}{4G_N}\int_{r_*}^{r_\infty}dr_0\ \frac{\sqrt{r_0^4-r_*^4}}{r_0^4}\left[2r_0^2h^{(2)}(t-1/r_0,r_0)-f^{(2)}(t-1/r_0,r_0)\right]\nonumber\\
&&+\epsilon^2\frac{\ell_\perp}{2G_N}r_2(\ell/2)+\mathcal{O}(\epsilon^4)\ .
\end{eqnarray}
Notice that the integrand in the first line is completely determined once the quench profile $J$ is specified, since the metric coefficients $f^{(2)}$ and $h^{(2)}$ are known from (\ref{eq:LifQuenchSolPart2}). The constant $r_*$ is related to the boundary separation $\ell$ via the analytic expression (\ref{eq:LifEEparameters})\footnote{Even without knowing the explicit solution (\ref{eq:LifEEsol}) to the equations of motion we still could find the boundary separation $\ell$ by simply looking at equation (\ref{eq:LifEEDerivativesrveqb}) as a differential equation for $x(r_0)$ instead of $r_0(x)$, then integrating from $x=0$ to $x=\ell/2$ and using the boundary conditions $r_0(0)=r_*,r_0(\ell/2)=r_\infty\rightarrow\infty$ to get $$\ell=2\int_{r_*}^\infty\frac{dr}{r^2\sqrt{\frac{r^4}{r_*^4}-1}}\ .$$}. As we shall see, the contribution of $r_2(\ell/2)$ in the second line will depend on time and therefore must be taken into account into the time evolution of $\delta S_A(t)$. 

However, there are two immediate problems with the expression (\ref{eq:LifEEFinal}): the integral in the first line of (\ref{eq:LifEEFinal}) diverges due to the contribution near the boundary $r=r_\infty\rightarrow\infty$, as well as the $r_2(\ell/2)$ term diverges due to our boundary conditions, and we need a regularization procedure in order to get a finite result for the entanglement entropy. In practice this can be done by using the large $r$ regulator $r_\infty$ to identify the divergences. Namely, our goal will be to split the entanglement entropy into two contributions: a divergent one, regulated by $r_\infty$, and a finite subleading one (which will be studied in detail), i.e.,
\begin{equation}\label{eq:LifEEDivAndFinite}
\delta S_A(t) = \delta S_{A_\textrm{div}}(t) + \delta S_{A_\textrm{finite}}(t).
\end{equation}
An alternative way would be to use the renormalized version of the entanglement entropy introduced in \cite{Liu:2012eea}, but we shall not pursue this here.

We first regulate the term $r_2(\ell/2)$. In order to find the radial profile correction $r_2(x)$ we need to solve the Euler-Lagrange equations for $r_2(x),v_2(x)$ appearing in the functional (\ref{eq:LifEEOrder2}). They consist of a complicated set of coupled differential equations involving the order $\epsilon^2$ metric coefficients $f^{(2)},h^{(2)}$ (and their derivatives) as well as the order $\epsilon^0$ profiles $r_0,v_0$ found before, which is hardly enlightening to show here. However, since we just need the value of $r_2$ at $x=\ell/2$ we can solve these equations only for $x$ near $\ell/2$, in which case they simplify considerably. The order $0$ radial profile appearing in (\ref{eq:LifEEsol}) in this regime takes the simple power-law form 
\begin{equation}
 r_0(x)=\left(\frac{r_*^2}{3y}\right)^{1/3}+\cdots\ ,
\end{equation}
where $y\equiv\ell/2-x\rightarrow0$. By inserting this result in the aforementioned pair of equations and solving perturbatively in $y$ it is easy to find the profile $r_2(x)$ as being
\begin{eqnarray}
 r_2(x) &=& \frac{1}{3}J(t)^2\ln\left(\frac{3y}{r_*^2}\right)\left(\frac{r_*^2}{3y}\right)^{1/3}+\frac{1}{12}J(t)\dot{J}(t)\left(15+4\ln\frac{r_*^2}{3y}\right)+\cdots\nonumber\\
 &=& - J(t)^2r_0(x)\ln r_0(x)+J(t)\dot{J}(t)\ln r_0(x)+\frac{5}{4}J(t)\dot{J}(t)+\cdots
\end{eqnarray}
for small $y$. The first two terms are clearly divergent for $x\rightarrow\ell/2$ ($y\rightarrow0$), while the terms in the ellipsis all vanish in this limit. Using the same regulator $r_\infty$ introduced before, i.e., $r_0(\ell/2)=r_\infty$, the value of $r_2$ at $x=\ell/2$ is then found to be
\begin{equation}\label{eq:LifEEr2}
 r_2(\ell/2) = - J(t)^2r_\infty\ln r_\infty+J(t)\dot{J}(t)\ln r_\infty+\frac{5}{4}J(t)\dot{J}(t)\ .
\end{equation}

Now we discuss the regularization of the integral term in (\ref{eq:LifEEFinal}). The divergent part comes from the leading behavior of the metric functions $f^{(2)}$ and $h^{(2)}$ near $r_0\rightarrow\infty$. It follows from expressions (\ref{eq:LifQuenchSolPart2}) that the large $r_0$ behavior of the combination $2r_0^2h^{(2)}-f^{(2)}$ appearing inside the integral is
$$2r_0^2h^{(2)}(t-1/r_0,r_0)-f^{(2)}(t-1/r_0,r_0)=\frac{1}{2}r_0^2J(t)^2+\cdots\ .$$
Therefore, in order to identify the divergences one just needs to plug this result into the integrand and evaluate the integral with the UV regulator $r_\infty$, namely
\begin{eqnarray}\label{eq:LifEEintdivterm}
 \delta S_{A_\textrm{int,div}}(t)&=&\epsilon^2\frac{\ell_\perp}{4G_N}\int_{r_*}^{r_\infty}dr_0\ \frac{\sqrt{r_0^4-r_*^4}}{r_0^4}\left[\frac{1}{2}r_0^2J(t)^2\right]\nonumber\\
 &=& \epsilon^2\frac{\ell_\perp}{4G_N}\left[\frac{1}{2}J(t)^2r_\infty\right]+\epsilon^2\frac{\ell_\perp}{4G_N}\left[\frac{\sqrt{\pi}\Gamma(-1/4)}{16\Gamma(5/4)}r_*J(t)^2\right]\ .
\end{eqnarray}

Therefore, the finite part of the entanglement entropy introduced in (\ref{eq:LifEEDivAndFinite}) follows simply from the general expression (\ref{eq:LifEEFinal}) by subtracting the divergent terms (all of them properly identified by the regulator $r_\infty$ in equations (\ref{eq:LifEEr2}),(\ref{eq:LifEEintdivterm})). The final result, written explicitly in terms of the quench profile instead of the metric functions $f^{(2)},h^{(2)}$, reads
\begin{eqnarray}\label{eq:LifEEFinalRenormalized}
 \delta S_{{}_{A_\textrm{finite}}}(t)&=&\epsilon^2\frac{\ell_\perp}{4G_N}\left\{\int_{r_*}^{\infty}\ dr\frac{\sqrt{r^4-r_*^4}}{r^2} \left[\frac{J(t-1/r)^2-J(t)^2}{2}+\frac{J(t-1/r)\dot{J}(t-1/r)}{r}\right.\right.\nonumber\\
&&\left.\left.+\frac{3 \dot{J}(t-1/r)^2}{4r^2}+\frac{I(t-1/r)}{r^3}\right] + \frac{\sqrt{\pi}\Gamma(-1/4)}{16\Gamma(5/4)}r_*J(t)^2+\frac{5}{2}J(t)\dot{J}(t)\right\},
\end{eqnarray}
where once again we stress that $r_*$ is related to the boundary separation $\ell$ via (\ref{eq:LifEEparameters})\footnote{Here we have used a trick in order to extract the finite contribution to the integral in (\ref{eq:LifEEFinal}): instead of calculating the full original integral and then subtracting the divergent piece $\epsilon^2\frac{\ell_\perp}{4G_N}[\frac{1}{2}J(t)^2r_\infty]$ obtained in (\ref{eq:LifEEintdivterm}), we equivalently subtract the whole integral in (\ref{eq:LifEEintdivterm}) and add back separately the constant term coming from the lower limit. In this way, the UV divergence of the integral is cancelled directly in the integrand even before integrating (which is convenient for numerical integration) at the cost of adding back by hand the extra term.}. Notice that the integrand naturally vanishes at large $r$ and hence the result of the integral is indeed finite. 

In the following we will make a detailed study of this quantity for the two quench profiles of interest as a probe of the quench dynamics. In doing so, it will be convenient to ignore the prefactor of $\ell_\perp/4G_N$ by defining the entanglement entropy density (times $4G_N$) $\delta s_{{}_{A_\textrm{finite}}}(t)\equiv\frac{4G_N}{\ell_\perp}\delta S_{{}_{A_\textrm{finite}}}(t)$.

\begin{figure}[htp]
    \centering
    \subfigure[$\ell=2$ fixed and $\delta t$ varying]{\includegraphics[width=7.0cm,height=6.0cm]{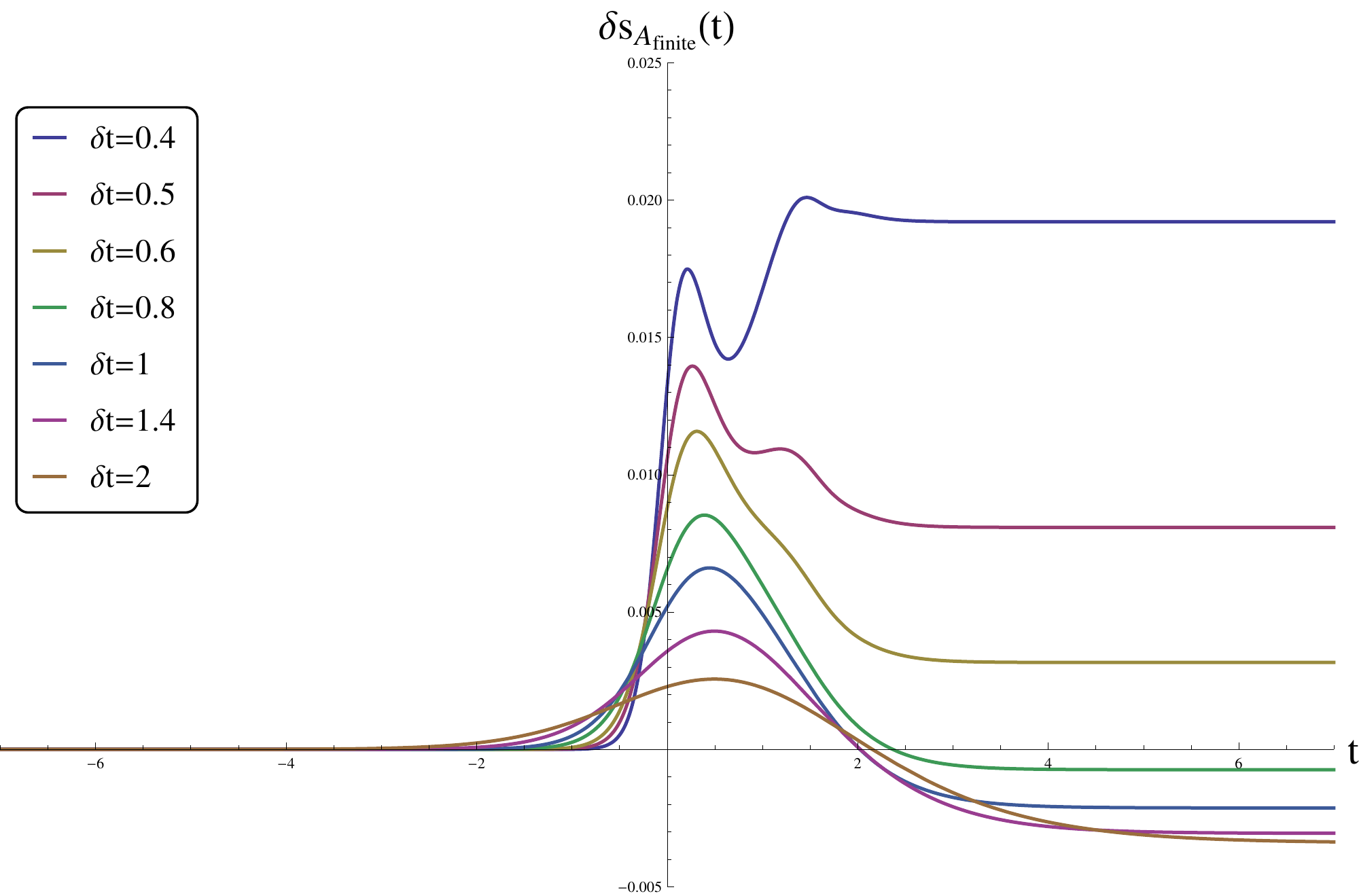}}\qquad
    \subfigure[$\delta t=1$ fixed and $\ell$ varying]{\includegraphics[width=7.0cm,height=6.0cm]{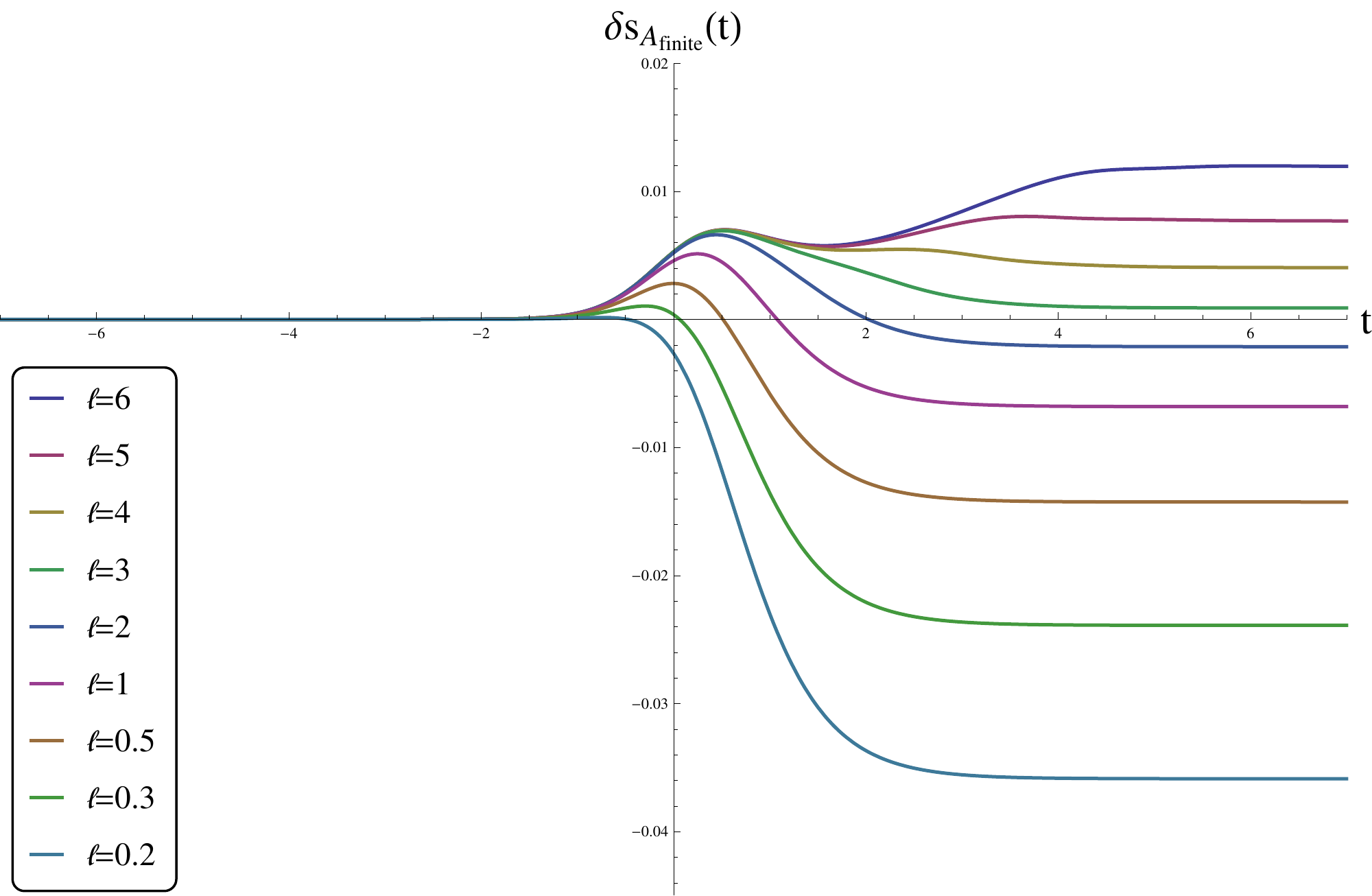}}\qquad
    \caption{\small Time evolution of $\delta s_{{}_{A_\textrm{finite}}}(t)\equiv\frac{4G_N}{\ell_\perp}\delta S_{{}_{A_\textrm{finite}}}(t)$ for the Tanh quench (\ref{eq:LifQuenchProfileTanh}). In (a) the boundary separation $\ell=2$ is fixed and we compare different quenching times $\delta t$. The curves go from $\delta t=0.4$ (top) to $2$ (bottom). For $\delta t\lesssim0.35$ the perturbative solution is expected to break down (for our choice of $\epsilon=0.1$), so we only show curves for $\delta t$ above this value. In (b) the quenching time $\delta t=1$ is fixed and we study the thermalization process at different length scales, from $\ell=0.2$ (bottom) to $\ell=6$ (top).}\label{fig:EETanh}
\end{figure}

In Figure \ref{fig:EETanh} we show the time evolution of the entanglement entropy for the $\tanh$ quench profile (\ref{eq:LifQuenchProfileTanh}). For simplicity we fix the value $\epsilon=0.1$, meaning that the final state Lifshitz theory will have the dynamical exponent $z=1+\epsilon^2=1.01$. 

In part (a) the value of the boundary separation is fixed to be $\ell=2$ so as to study the effect of the quenching time $\delta t$. We recall from the discussion above that the minimal surface penetrates inside the bulk from $r=\infty$ up to $r_*(\ell)$ given by equation (\ref{eq:LifEEparameters}), which in the case of $\ell=2$ corresponds to $r_*=0.599$. This means that one can trust our solution to calculate the entropy with such a value of $\ell$ as long as the final state Lifshitz black brane forms at $r_h$ sufficiently away from this value\footnote{We will adopt in this work the convention of $r_*>2r_h$ for what we mean by \lq\lq sufficiently away\rq\rq. Thus, in the present case, for example, we demand $r_h\lesssim0.3$.}. It follows from the definition of $r_h$ that this constrains the quenching time to be $\delta t>0.37$ (of course this constraint will change for a different $\epsilon$), and for that reason we show in the plot a comparison of many curves with different values of $\delta t$ only above this value. It can be seen from the plot that despite the quench $J(t)$ being a monotonically increasing function, the time evolution of the (finite part of) entanglement entropy is never monotonic and differs qualitatively depending on the quenching rate $\delta t$. Namely, fast enough quenches induce an oscillatory behavior at intermediate times before the thermal state is reached, while slower quenches do not. Increasing the value of $\delta t$ we see that the equilibration curves become smoother, approaching the adiabatic regime studied in \cite{Basu:2011ft}. Remarkably, by comparing the equilibrium value of the entanglement entropy at the Lifshitz point with the initial background value we see that there may be an increase or decrease depending on the quenching time: slow quenches ($\delta t\gtrsim0.8$) cause an entanglement loss in the process, while for quenches faster than these the amount of entanglement entropy is increased (the faster the quench is, the bigger the gap between the final and initial values becomes).

In part (b) we now fix the quenching time to be $\delta t=1$ and analyze the thermalization curves for different boundary separations $\ell$ (i.e., at different energy scales in the boundary gauge theory). Note that with $\epsilon=0.1$ and $\delta t=1$ the horizon radius of the final state black hole is fixed at $r_h=0.11$, so one can trust the calculation for all length scales up to $\ell\simeq6$ (for which $r_*\sim2r_h$).
It is clear from the figure that the thermalization of the entanglement entropy is a top-down process, i.e., short-scale entanglement entropy equilibrates before its large-distance counterpart. From the dual gauge theory point of view, this result once again suggests that the dynamical breaking of the relativistic scaling symmetry to a Lifshitz symmetry happens faster at short distances (high energies).
Another interesting aspect noted from the plot is that the dynamics (as told from the entanglement entropy) is qualitatively different at distinct length scales on the boundary. Namely, at very small distances ($\ell\sim0.2$) the entropy decreases monotonically in the whole process towards its final value (which is considerably less than the initial one). At larger distances, on the other hand, the dynamics becomes non-monotonic, the gap between the final and initial values is decreased, and the value of the entanglement entropy at the Lifshitz point can be even greater than the background one (for $\ell\gtrsim3$).

\begin{figure}[htp]
    \centering
    \subfigure[$\ell=2$ fixed and $\delta t$ varying]{\includegraphics[width=7.0cm,height=6.0cm]{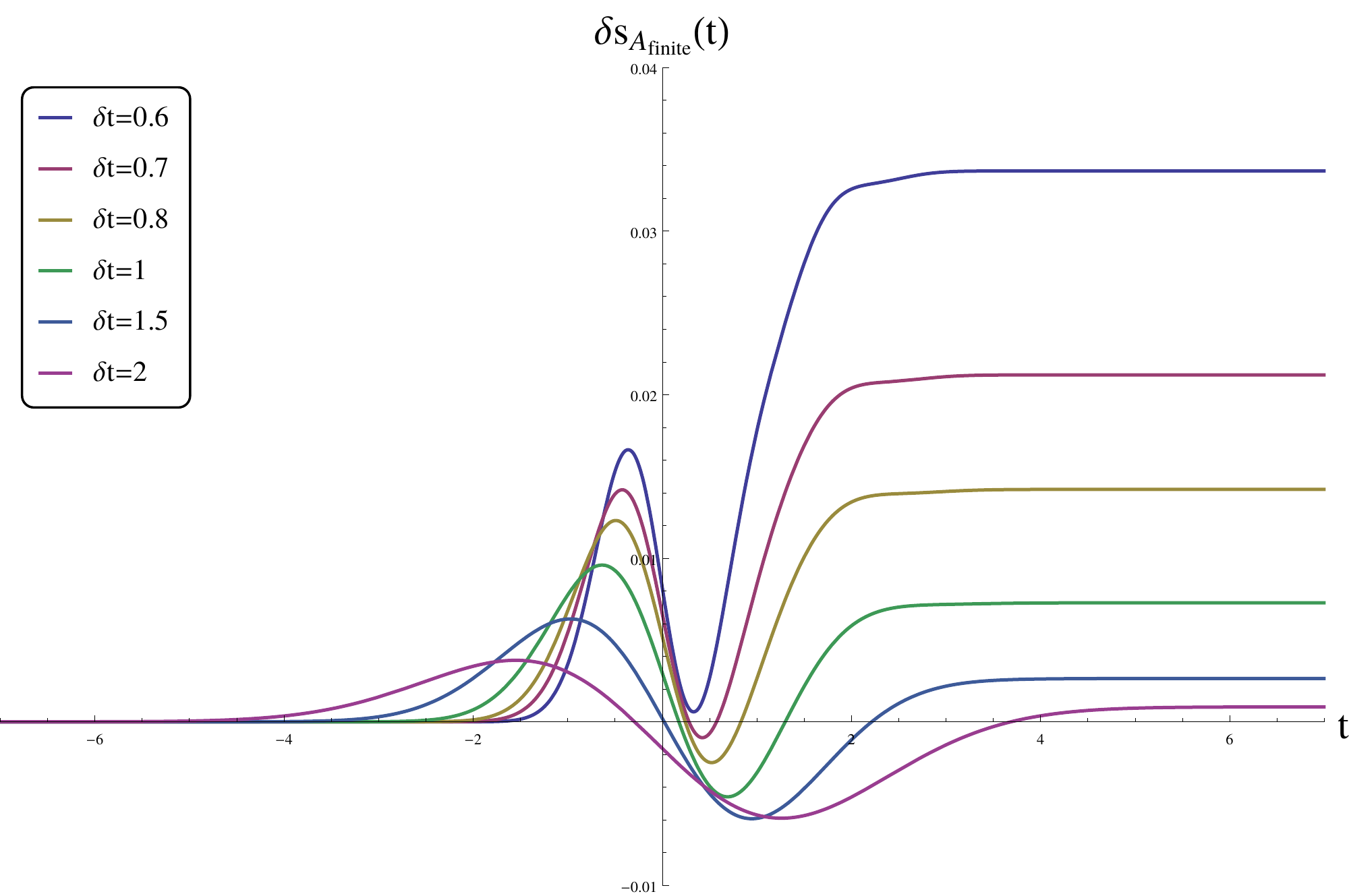}}\qquad
    \subfigure[$\delta t=1$ fixed and $\ell$ varying]{\includegraphics[width=7.0cm,height=6.0cm]{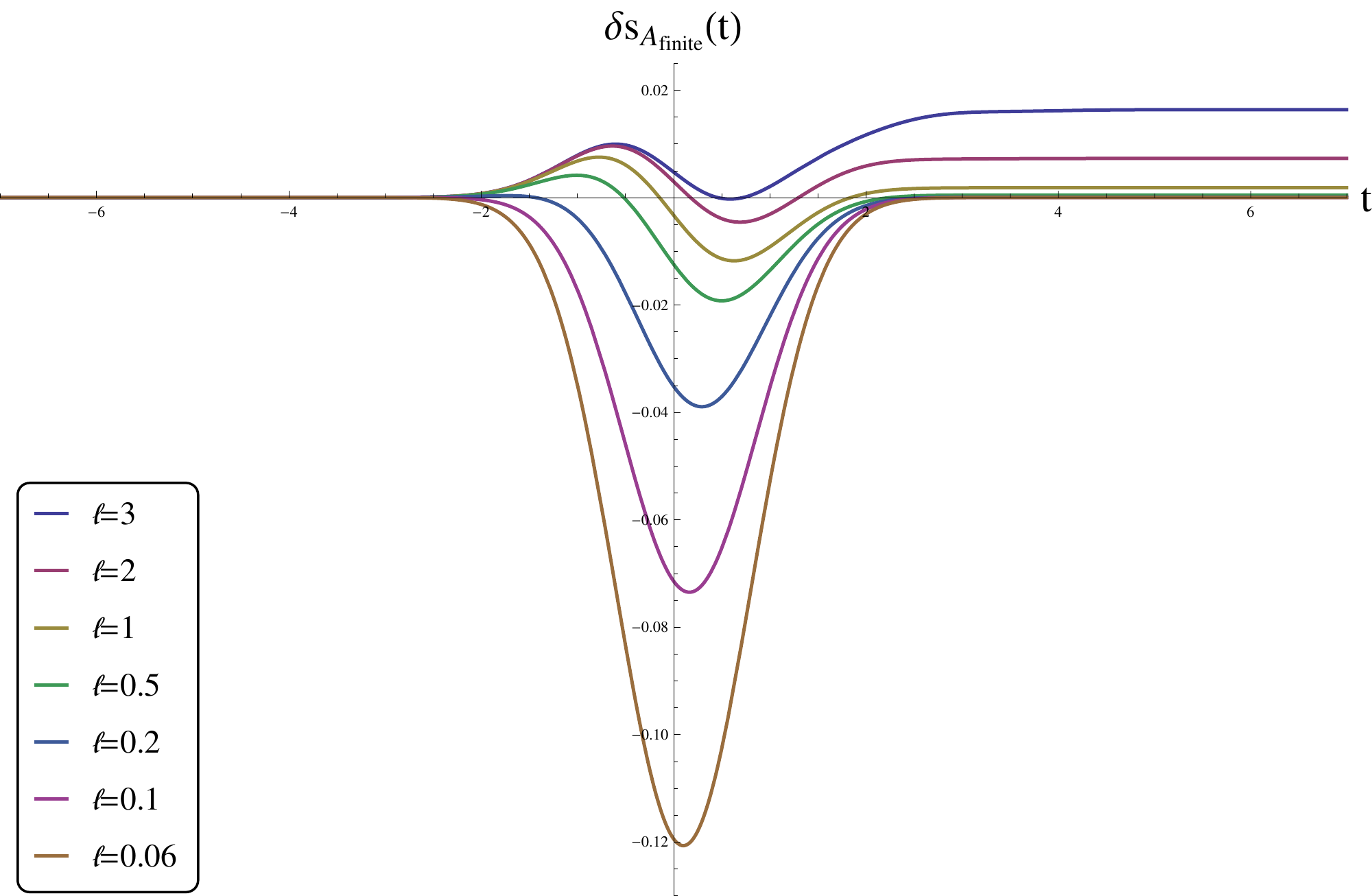}}\qquad
    \caption{\small Time evolution of $\delta s_{{}_{A_\textrm{finite}}}(t)\equiv\frac{4G_N}{\ell_\perp}\delta S_{{}_{A_\textrm{ren}}}(t)$ for the Gaussian quench (\ref{eq:LifQuenchProfileGauss}). In (a) the boundary separation $\ell=2$ is fixed and we compare different quenching times $\delta t$. The curves go from $\delta t=0.6$ (top) to $2$ (bottom). For $\delta t\lesssim0.6$ the perturbative solution is expected to break down (for our choice of $\epsilon=0.1$), so we only show curves for $\delta t$ above this value. In (b) the quenching time $\delta t=1$ is fixed and we study the thermalization process at different length scales, from $\ell=0.06$ (bottom) to $\ell=3$ (top).}\label{fig:EEGauss}
\end{figure}

In Figure \ref{fig:EEGauss} we make a similar analysis for the Gaussian quench profile (\ref{eq:LifQuenchProfileGauss}). Again we use the value $\epsilon=0.1$, but it should be kept in mind that now this does not correspond to the Lifshitz exponent since in the final state we have an asymptotically AdS black hole. In part (a) the value of the boundary separation is fixed ($\ell=2$) and the quenching time $\delta t$ is varied. The regime of validity of our solution now constrains $\delta t\gtrsim0.6$, which is the reason why we show only curves with $\delta t$ above this value. We notice from the plot that the time evolution is always non-monotonic, as in the case of Tanh profile analyzed above, but the form of the curves is slightly different. The breaking of the relativistic scaling at intermediate times and its subsequent restoration manifests here as an oscillatory behavior of the entanglement entropy before reaching the final value.  An important difference with respect to the Tanh quench previously analyzed is that the final equilibrium value of the entropy is always bigger than the initial one, i.e., there is always an entanglement growth in the process regardless of the value of $\delta t$. The quenching time sets the gap between the final and initial values for the entanglement entropy, namely, the gap is larger for faster quenches.

In part (b) it is the quenching time that is fixed to $\delta t=1$, and we analyze the thermalization curves at different values of $\ell$. Note that by choosing $\epsilon=0.1$ and $\delta t=1$ the horizon radius of the final state black hole is now fixed at $r_h=(3\sqrt{\pi}\epsilon^2/8\delta t^3)^{1/3}=0.19$, so one can trust the calculation for all length scales up to $\ell\simeq3$ (for which $r_*=0.40>2r_h$). We see from the figure the same top-down thermalization observed for the Tanh profile. It is also interesting to notice that at distance scales up to $\ell\simeq0.5$ the gap between the final and initial values for the entanglement entropy is almost zero. As already discussed, this should not be a surprise since differences between the final state AdS black hole and the pure AdS initial state can only be seen if we probe deep inside the bulk (i.e., for minimal surfaces with large $\ell$). In the case of the Tanh quench, where we had a Lishitz black brane in the final state, such a gap had no reason to be small due to the $\ln r$ term in the metric (\ref{eq:LifFinalStateBB}) which can be sensed even without penetrating deep into the bulk. 

We close by noticing from Figures \ref{fig:EETanh} and \ref{fig:EEGauss} that $\delta s_{{}_{A_\textrm{finite}}}$ eventually becomes negative for different combinations of $\ell$ and $\delta t$. However, this is not a problem since it represents only the finite contribution to the entanglement entropy. The full entanglement entropy has an additional divergent contribution (regulated by the cutoff $r_\infty$) in such a way that it always increases.

\section{Conclusions}\label{Conclusions}

We have considered here the problem of holographic quenches leading to a breaking of the standard relativistic scaling symmetry towards a Lifshitz scaling with $z=1+\epsilon^2$ ($\epsilon\ll1$). The quenching operator is (the time component of) a vector operator $\mathcal{V}_t$ with dimension $\Delta=d$, in which case the Lifshitz theory can be understood as a standard deformation of the CFT.

After introducing the perturbative setup in the bulk, we have found (to order $\epsilon^2$) the gravity solution describing the quench dynamics and discussed its regime of validity. In particular, this regime excludes the case of infinitely fast quenches. The solution interpolates between pure AdS space at past infinity and an asymptotically Lifshitz black hole at future infinity. This means that the corresponding non-relativistic dual field theory appearing at the end state is always at finite temperature or, conversely, that it is impossible to reach the vacuum state of the Lifshitz theory from the CFT vacuum using the continuous quench mechanism proposed here. 

We have also probed the nonequilibrium dynamics following the breaking of the relativistic scaling using both local (1-point correlators of operators in the boundary theory) and non-local (the entanglement entropy) observables, as well as the apparent and event horizons. Both horizons were shown to be monotonically increasing functions of time, with the apparent horizon being inside the event horizon during the whole process, agreeing with what is expected for physically reasonable collapse processes. The full time evolution of 1-point functions of $T_{ab}$ and the quenching operator $\mathcal{V}^a$ in the boundary was obtained analytically, and they were shown to satisfy all the expected Ward identities. However, being local observables, they are not sensitive to physics at different scales. 
Using the entanglement entropy the thermalization process was probed at different length scales $\ell$ in the boundary theory and for different values of the quenching rate $\delta t$. Specifically, we have concluded that the equilibration is a top-down process, i.e., the symmetry breaking takes place faster for UV modes than for low energy modes. In addition, the curves are slightly different depending on the value of $\delta t$ and the gap between the final and initial values increases for faster quenches.

The present work can be generalized in many ways, such as changing the number of dimensions (we used $d=3$ for the boundary theory) or the quench profile. More interesting generalizations to pursue are the inclusion of a hyperscaling violation parameter or the study of quenches in the Schr\"odinger background (in this case reference \cite{Guica:2010sw} may be helpful), which we leave for a future work.

\acknowledgments

We would like to thank Diego Trancanelli, Andrea Prudenziati, Olivera Mi\u skovi\u c, Rodrigo Olea, and Alan Pavan for valuable discussions and suggestions. We also thank the anonymous referee for enlightening suggestions. G. Camilo and E. Abdalla thank CNPq for financial support. E. Abdalla also thanks the support from FAPESP.

\appendix
\section{Holographic renormalization and 1-point functions}\label{LifHoloRenorm}

\subsection{General results for an arbitrary solution}

The holographic renormalization of the Einstein-Proca model (\ref{eq:LifshitzMassiveVectorAction}) as well as the calculation of renormalized 1-point functions has been carried out in full detail in \cite{Korovin:2013bua} (to order $\epsilon^2$). Here we summarize the main results, which hold for an arbitrary solution to the bulk equations of motion in $3+1$ dimensions, before particularizing to our case of interest. The reader is referred to \cite{Korovin:2013bua} for the details.

As usual in holographic renormalization, the calculation is done using Fefferman-Graham (FG) coordinates $(t,\rho)$ rather than the EF coordinates $(v,r)$ used in Section \ref{Quenches}. The main reason is that in the latter the \lq\lq radial\rq\rq~direction $r$ is not orthogonal to the spacetime boundary located at infinity, while the FG coordinate $\rho$ is spacelike and, hence, one can choose a timelike planar cutoff surface by simply setting $\rho=\rho_\infty$. Namely, the metric and vector field are parametrized in FG coordinates as
\begin{subequations}\label{eq:LifFeffermanGraham}
\begin{align}
ds^2 &= \frac{d\rho^2}{\rho^2} + \rho^2g_{ab}dx^adx^b\\ 
A_\mu &=A_\rho d\rho+A_a dx^a\ ,
\end{align}
\end{subequations}
where $x^a\equiv(t,x^i)$ are boundary coordinates and the $\epsilon$ expansion is taken as before,
\begin{subequations}\label{eq:LifFeffermanGrahamEpsilon}
\begin{align}
g_{ab}(\rho,x;\epsilon) &= g^{(0)}_{ab}(\rho,x)+\epsilon^2g^{(2)}_{ab}(\rho,x)+\mathcal{O}(\epsilon^4)\\
A_{a}(\rho,x;\epsilon) &= \epsilon\rho A^{(1)}_{a}(\rho,x)+\mathcal{O}(\epsilon^3)\\
A_{\rho}(\rho,x;\epsilon) &= \epsilon\rho A^{(1)}_{\rho}(\rho,x)+\mathcal{O}(\epsilon^3)\ .
\end{align}
\end{subequations}
Since the divergences in the on-shell action occur only due to the contribution at the boundary, just the large $\rho$ behavior of the quantities is necessary. Thus, in addition to the $\epsilon$ expansion, each function appearing above admits also an asymptotic expansion\footnote{Whose order in $\frac{1}{\rho}$ we shall denote by a square bracket subscript $_{[n]}$.} near the boundary of the form
\begin{subequations}\label{eq:LifFeffermanGrahamAsymptotic}
\begin{align}
 g^{(0)}_{ab}(\rho,x) &= g^{(0)}_{[0]ab}(x)+\frac{1}{\rho^2}g^{(0)}_{[2]ab}(x)+\frac{1}{\rho^3}g^{(0)}_{[3]ab}(x)+\cdots\\
 A^{(1)}_{a}(\rho,x) &= A^{(1)}_{[0]a}(x)+\frac{1}{\rho^2}A^{(1)}_{[2]a}(x)+\frac{1}{\rho^3}\left(A^{(1)}_{[3]a}(x)+\tilde{A}^{(1)}_{[3]a}(x)\ln\rho\right)+\cdots\\
 A^{(1)}_{\rho}(\rho,x) &= A^{(1)}_{[0]\rho}(x)+\frac{1}{\rho^2}A^{(1)}_{[2]\rho}(x)+\frac{1}{\rho^3}\left(A^{(1)}_{[3]\rho}(x)+\tilde{A}^{(1)}_{[3]\rho}(x)\ln\rho\right)+\cdots\\
 g^{(2)}_{ab}(\rho,x) &= h^{(2)}_{[0]ab}(x)\ln\rho+\frac{\left(g^{(2)}_{[2]ab}(x)+h^{(2)}_{[2]ab}(x)\ln\rho\right)}{\rho^2}+\frac{\left(g^{(2)}_{[3]ab}(x)+h^{(2)}_{[3]ab}(x)\ln\rho\right)}{\rho^3}+\cdots\ .
\end{align}
\end{subequations}

With the FG expansion above the equations of motion can be solved order by order in $\epsilon$ to yield the most general asymptotic solution for the metric and vector field given the non-normalizable modes $g^{(0)}_{[0]ab}(x)$ and $A^{(1)}_{[0]a}(x)$ as arbitrary Dirichlet data on the boundary. This asymptotic solution is used to calculate the regulated on-shell action and identify the divergent contributions, which are then cancelled by appropriate counterterms defined at the regulated boundary $\rho=\rho_\infty$. The resulting renormalized on-shell action (to order $\epsilon^2$) is
\begin{eqnarray}\label{eq:LifshitzRenormalizedAction}
S_\textrm{ren} &=& S_\textrm{on-shell} + S_{_\textrm{GH}} + S^{(0)}_\textrm{ct} + S^{(2)}_\textrm{ct}\nonumber\\
&=& \frac{1}{16\pi G}\int d^4x\sqrt{-g}\left[-6-\frac{1}{4}F_{\mu\nu}F^{\mu\nu}\right]+\frac{1}{8\pi G}\int_\partial d^3x\sqrt{-\gamma}K+\nonumber\\
&&-\frac{1}{16\pi G}\int_\partial d^3x\sqrt{-\gamma}\left(4+R[\gamma]\right)+\frac{1}{32\pi G}\int_\partial d^3x\sqrt{-\gamma}A_aA^a+\cdots\ ,
\end{eqnarray}
where $S_{_\textrm{GH}}$ is the usual Gibbons-Hawking boundary term, $S^{(0)}_\textrm{ct}$ is the order $\epsilon^0$ (pure gravity) counterterm obtained in \cite{deHaro:2000vlm}
and $S^{(2)}_\textrm{ct}$ is the counterterm needed to cancel the leading divergences\footnote{There are also subleading divergences at order $\epsilon^2$ which require additional counterterms to be removed, but such extra pieces do not contribute to the 1-point functions and can be ignored for our purposes.} at order $\epsilon^2$. 

The desired correlation functions then follow simply from functional differentiation of $S_\textrm{ren}$ with respect to the sources, i.e.,
\begin{equation}
 \delta S_\textrm{ren}\big[g^{{(0)}}_{[0]ab},A^{(1)}_{[0]a}\big] = -\int d^3x\sqrt{-g^{_{(0)}}_{[0]}}\left[\frac{1}{2}\langle T_{ab}\rangle\delta g^{(0)ab}_{[0]}+\langle \mathcal{V}^a\rangle\delta A^{(1)}_{[0]a}\right]\ ,
\end{equation}
the result being
\begin{subequations}\label{eq:LifCorrelatorsGeneral}
 \begin{align}
 \big\langle \mathcal{V}^{a}\big\rangle &= \epsilon\left[\frac{1}{16\pi G}g^{(0)ab}_{[3]} A^{(1)}_{[0]b} - \frac{3}{16\pi G}A^{(1)a}_{[3]}\right]+\mathcal{O}(\epsilon^3)\\
 \big\langle T_{ab}\big\rangle &= \frac{3}{16\pi G}g^{(0)}_{[3]ab} -\frac{\epsilon^2}{16\pi G}\left[h^{(2)}_{[3]ab}-3g^{(2)}_{[3]ab} - \frac{1}{4}A^{(1)}_{[0]c}A^{(1)c}_{[0]}g^{(0)}_{[3]ab}-A^{(1)c}_{[0]}g^{(0)}_{[3]cd}A^{(1)d}_{[0]}g^{(0)}_{[0]ab}\right.\nonumber\\
 &\left. + A^{(1)}_{[0]c}A^{(1)c}_{[3]}g^{(0)}_{[0]ab}+A^{(1)}_{[0]a}A^{(1)}_{[3]b}+A^{(1)}_{[0]b}A^{(1)}_{[3]a}\right] + \mathcal{O}(\epsilon^4)\ .
 \end{align}
\end{subequations}

\subsection{Particularizing to the quench solution}

The only thing needed to make contact between our case and the general results presented above is to express our solution (\ref{eq:LifQuenchSolPart1})-(\ref{eq:LifQuenchSolPart2}) in the FG form given in (\ref{eq:LifFeffermanGraham})-(\ref{eq:LifFeffermanGrahamAsymptotic}). This is done by equating the two line elements and writing the EF coordinates $v$ and $r$ as functions of the new FG coordinates $(t,\rho)$, which provides a set of 3 equations to be solved for $v(t,\rho),r(t,\rho)$ and the metric component $g_{tt}(t,\rho)$. Namely,
\begin{subequations}\label{eq:LifFeffermanGrahamTransfEqs}
 \begin{align}
2hr'v'-fv'^2 &= \frac{1}{\rho^2}\\
2h\dot{r}\dot{v}-f\dot{v}^2 &= \rho^2g_{tt}\\
h(r'\dot{v}+\dot{r}v')-fv'\dot{v} &= 0\ ,
 \end{align}
\end{subequations}
with $h(v,r)\equiv1+\epsilon^2h^{(2)}(v,r)$, $f(v,r)\equiv r^2+\epsilon^2f^{(2)}(v,r)$ and a prime (dot) denotes $\partial_\rho$ ($\partial_t$).
From the above one can express $g_{tt}$ in terms of $v(t,\rho)$ alone, whilst the spatial components $g_{ij}$ depend only on $r(t,\rho)$ as follows straightforwardly from the definition (\ref{eq:LifFeffermanGraham}), namely 
\begin{equation}\label{eq:LifFeffermanGrahamTransfMetric}
 g_{tt}=-\frac{\dot{v}(t,\rho)^2}{\rho^4 v'(t,\rho)^2}\ ,\qquad\qquad g_{ij}(t,\rho)=\frac{r(t,\rho)^2}{\rho^2}\delta_{ij}\ .
\end{equation}

Equations (\ref{eq:LifFeffermanGrahamTransfEqs}) can be solved order by order in $\epsilon$ by writing
\begin{subequations}
 \begin{align}
v(t,\rho) &= v^{(0)}(t,\rho) + \epsilon^2 v^{(2)}(t,\rho) + \mathcal{O}(\epsilon^4)\\
r(t,\rho) &= r^{(0)}(t,\rho) + \epsilon^2 r^{(2)}(t,\rho) + \mathcal{O}(\epsilon^4)\ .
 \end{align}
\end{subequations}
Actually to order $\epsilon^0$ the coordinate transformation is already known: since the bulk solution in this case is simply pure AdS space, the Poincar\'e coordinates (\ref{eq:StaticLifSolution}) do the job as our FG coordinates, i.e., 
\begin{equation*}
 v^{(0)}(t,\rho)=t-\frac{1}{\rho}\ ,\qquad\qquad r^{(0)}(t,\rho)=\rho\ .
\end{equation*}
The solution to order $\epsilon^2$ can be found in the large $\rho$ asymptotic expansion using a power series ansatz with log terms of the form 
\begin{equation*}
 v^{(2)}(t,\rho) = \sum_{n}(c_n(t)+\tilde{c}_n(t)\ln\rho)\rho^n
\end{equation*}
(and similarly for $r^{(2)}$). The complete resulting coordinate transformation is given by
\begin{subequations}
 \begin{align}
 v(t,\rho) &= t-\frac{1}{\rho} + \epsilon^2 \left[\frac{1}{4}\int_{-\infty}^t J(u)^2du+\frac{3J(t)^2 \ln\rho+2J(t)^2}{4\rho}-\frac{6J(t)\dot{J}(t)\ln\rho+7J(t)\dot{J}(t)}{8\rho ^2}\right.\nonumber\\
 &\quad\left.+\frac{48\left(J(t) \ddot{J}(t)+\dot{J}(t)^2\right)\ln\rho+5\left(14 J(t)\ddot{J}(t)+11 \dot{J}(t)^2\right)}{144\rho^3}+\cdots\right] + \mathcal{O}(\epsilon^4)\\
 r(t,\rho) &= \rho +\epsilon^2 \left[-\frac{1}{4}J(t)^2 \rho\ln\rho+\frac{\dot{J}(t)^2-2J(t)\ddot{J}(t)}{16\rho} + \frac{6I(t)+2J(t)\dddot{J}(t)-3\dot{J}(t)\ddot{J}(t)}{36\rho ^2}\right.\nonumber\\
 &\quad\left.+ \frac{-8\dot{I}(t)-J(t)\ddddot{J}(t)+3\ddot{J}(t)^2+2\dddot{J}(t)\dot{J}(t)}{64\rho^3}+\cdots\right] + \mathcal{O}(\epsilon^4)\ .
 \end{align}
\end{subequations}

As a consequence, using (\ref{eq:LifFeffermanGrahamTransfMetric}) our metric and vector field (\ref{eq:LifQuenchSolPart1})-(\ref{eq:LifQuenchSolPart2}) can be cast in the desired FG form (\ref{eq:LifFeffermanGraham})-(\ref{eq:LifFeffermanGrahamAsymptotic}) with the following non-vanishing coefficient functions
\begin{subequations}\label{eq:LifFeffermanGrahamAsymptoticQuench}
 \begin{align}
 g^{(0)}_{[0]ab}(x) &= \eta_{ab}\\
 A^{(1)}_{[0]t}(x) &= \sqrt{2}J(t),\qquad A^{(1)}_{[3]t}(x)=-\frac{\dddot{J}(t)}{3\sqrt{2}},\qquad  A^{(1)}_{[3]\rho}(x)=\frac{\dot{J}(t)}{\sqrt{2}}\\
 h^{(2)}_{[0]tt}(x) &= -\frac{3}{2} J(t)^2,\qquad h^{(2)}_{[2]tt}(x)=-\frac{J(t)\ddot{J}(t)+\dot{J}(t)^2}{2}\\
 g^{(2)}_{[2]tt}(x) &= \frac{\dot{J}(t)^2-4J(t)\ddot{J}(t)}{8},\qquad g^{(2)}_{[3]tt}(x)=\frac{2I(t)-\dot{J}(t)\ddot{J}(t)}{3}\\
 h^{(2)}_{[0]ij}(x) &= -\frac{1}{2} J(t)^2\delta_{ij}\\
 g^{(2)}_{[2]ij}(x) &= \frac{\dot{J}(t)^2-2J(t)\ddot{J}(t)}{8}\delta_{ij},\quad g^{(2)}_{[3]ij}(x)=\frac{6I(t)+2 J(t)\dddot{J}(t)-3\dot{J}(t)\ddot{J}(t)}{18}\delta_{ij}\ .
 \end{align}
\end{subequations}
Finally, plugging the identifications (\ref{eq:LifFeffermanGrahamAsymptoticQuench}) into the general expressions (\ref{eq:LifCorrelatorsGeneral}) results in the correlators presented in (\ref{eq:LifCorrelatorsQuench}).

\bibliographystyle{JHEP}
\bibliography{myref}

\providecommand{\href}[2]{#2}\begingroup\raggedright\begin{thebibliography}{10}

\bibitem{2010LNP...802...21M}
S.~Mondal, D.~Sen, and K.~Sengupta, {\it {Non-equilibrium Dynamics of Quantum
  Systems: Order Parameter Evolution, Defect Generation, and Qubit Transfer}},
  {\em Lecture Notes in Physics} {\bf 802} (2010)
  [\href{http://arxiv.org/abs/0908.2922}{{\tt arXiv:0908.2922}}].

\bibitem{2011RvMP...83..863P}
A.~Polkovnikov, K.~Sengupta, A.~Silva, and M.~Vengalattore, {\it {Colloquium:
  Nonequilibrium dynamics of closed interacting quantum systems}},  {\em
  Reviews of Modern Physics} {\bf 83} (2011) 863--883,
  [\href{http://arxiv.org/abs/1007.5331}{{\tt arXiv:1007.5331}}].

\bibitem{Calabrese:2005in}
P.~Calabrese and J.~L. Cardy, {\it {Evolution of entanglement entropy in
  one-dimensional systems}},  {\em J. Stat. Mech.} {\bf 0504} (2005) P04010,
  [\href{http://arxiv.org/abs/cond-mat/0503393}{{\tt cond-mat/0503393}}].

\bibitem{Calabrese:2006rx}
P.~Calabrese and J.~L. Cardy, {\it {Time-dependence of correlation functions
  following a quantum quench}},  {\em Phys. Rev. Lett.} {\bf 96} (2006) 136801,
  [\href{http://arxiv.org/abs/cond-mat/0601225}{{\tt cond-mat/0601225}}].

\bibitem{2010PhRvB..81a2303D}
C.~de~Grandi, V.~Gritsev, and A.~Polkovnikov, {\it {Quench dynamics near a
  quantum critical point}},  {\em Phys. Rev. B} {\bf 81} (2010) 012303,
  [\href{http://arxiv.org/abs/0909.5181}{{\tt arXiv:0909.5181}}].

\bibitem{Calabrese:2007rg}
P.~Calabrese and J.~Cardy, {\it {Quantum Quenches in Extended Systems}},  {\em
  J. Stat. Mech.} {\bf 0706} (2007) P06008,
  [\href{http://arxiv.org/abs/0704.1880}{{\tt arXiv:0704.1880}}].

\bibitem{2010AdPhy..59.1063D}
J.~Dziarmaga, {\it {Dynamics of a quantum phase transition and relaxation to a
  steady state}},  {\em Advances in Physics} {\bf 59} (2010) 1063--1189,
  [\href{http://arxiv.org/abs/0912.4034}{{\tt arXiv:0912.4034}}].

\bibitem{Maldacena:1997re}
J.~M. Maldacena, {\it {The Large N limit of superconformal field theories and
  supergravity}},  {\em Int. J. Theor. Phys.} {\bf 38} (1999) 1113--1133,
  [\href{http://arxiv.org/abs/hep-th/9711200}{{\tt hep-th/9711200}}]. [Adv.
  Theor. Math. Phys.2,231(1998)].

\bibitem{Gubser:1998bc}
S.~S. Gubser, I.~R. Klebanov, and A.~M. Polyakov, {\it {Gauge theory
  correlators from noncritical string theory}},  {\em Phys. Lett.} {\bf B428}
  (1998) 105--114, [\href{http://arxiv.org/abs/hep-th/9802109}{{\tt
  hep-th/9802109}}].

\bibitem{Witten:1998qj}
E.~Witten, {\it {Anti-de Sitter space and holography}},  {\em Adv. Theor. Math.
  Phys.} {\bf 2} (1998) 253--291,
  [\href{http://arxiv.org/abs/hep-th/9802150}{{\tt hep-th/9802150}}].

\bibitem{Adams:2012th}
A.~Adams, L.~D. Carr, T.~Schäfer, P.~Steinberg, and J.~E. Thomas, {\it
  {Strongly Correlated Quantum Fluids: Ultracold Quantum Gases, Quantum
  Chromodynamic Plasmas, and Holographic Duality}},  {\em New J. Phys.} {\bf
  14} (2012) 115009, [\href{http://arxiv.org/abs/1205.5180}{{\tt
  arXiv:1205.5180}}].

\bibitem{Albash:2010mv}
T.~Albash and C.~V. Johnson, {\it {Evolution of Holographic Entanglement
  Entropy after Thermal and Electromagnetic Quenches}},  {\em New J. Phys.}
  {\bf 13} (2011) 045017, [\href{http://arxiv.org/abs/1008.3027}{{\tt
  arXiv:1008.3027}}].

\bibitem{Aparicio:2011zy}
J.~Aparicio and E.~Lopez, {\it {Evolution of Two-Point Functions from
  Holography}},  {\em JHEP} {\bf 12} (2011) 082,
  [\href{http://arxiv.org/abs/1109.3571}{{\tt arXiv:1109.3571}}].

\bibitem{Basu:2011ft}
P.~Basu and S.~R. Das, {\it {Quantum Quench across a Holographic Critical
  Point}},  {\em JHEP} {\bf 01} (2012) 103,
  [\href{http://arxiv.org/abs/1109.3909}{{\tt arXiv:1109.3909}}].

\bibitem{Basu:2012gg}
P.~Basu, D.~Das, S.~R. Das, and T.~Nishioka, {\it {Quantum Quench Across a Zero
  Temperature Holographic Superfluid Transition}},  {\em JHEP} {\bf 03} (2013)
  146, [\href{http://arxiv.org/abs/1211.7076}{{\tt arXiv:1211.7076}}].

\bibitem{Buchel:2012gw}
A.~Buchel, L.~Lehner, and R.~C. Myers, {\it {Thermal quenches in N=2*
  plasmas}},  {\em JHEP} {\bf 08} (2012) 049,
  [\href{http://arxiv.org/abs/1206.6785}{{\tt arXiv:1206.6785}}].

\bibitem{Liu:2013qca}
H.~Liu and S.~J. Suh, {\it {Entanglement growth during thermalization in
  holographic systems}},  {\em Phys. Rev.} {\bf D89} (2014), no.~6 066012,
  [\href{http://arxiv.org/abs/1311.1200}{{\tt arXiv:1311.1200}}].

\bibitem{Callebaut:2014tva}
N.~Callebaut, B.~Craps, F.~Galli, D.~C. Thompson, J.~Vanhoof, J.~Zaanen, and
  H.-b. Zhang, {\it {Holographic Quenches and Fermionic Spectral Functions}},
  {\em JHEP} {\bf 10} (2014) 172, [\href{http://arxiv.org/abs/1407.5975}{{\tt
  arXiv:1407.5975}}].

\bibitem{Rangamani:2015sha}
M.~Rangamani, M.~Rozali, and A.~Wong, {\it {Driven Holographic CFTs}},  {\em
  JHEP} {\bf 04} (2015) 093, [\href{http://arxiv.org/abs/1502.0572}{{\tt
  arXiv:1502.0572}}].

\bibitem{Keranen:2011xs}
V.~Keranen, E.~Keski-Vakkuri, and L.~Thorlacius, {\it {Thermalization and
  entanglement following a non-relativistic holographic quench}},  {\em Phys.
  Rev.} {\bf D85} (2012) 026005, [\href{http://arxiv.org/abs/1110.5035}{{\tt
  arXiv:1110.5035}}].

\bibitem{Sonner:2014tca}
J.~Sonner, A.~del Campo, and W.~H. Zurek, {\it {Universal far-from-equilibrium
  Dynamics of a Holographic Superconductor}},
  \href{http://arxiv.org/abs/1406.2329}{{\tt arXiv:1406.2329}}.

\bibitem{Das:2014jna}
S.~R. Das, D.~A. Galante, and R.~C. Myers, {\it {Universal scaling in fast
  quantum quenches in conformal field theories}},  {\em Phys. Rev. Lett.} {\bf
  112} (2014) 171601, [\href{http://arxiv.org/abs/1401.0560}{{\tt
  arXiv:1401.0560}}].

\bibitem{Das:2014hqa}
S.~R. Das, D.~A. Galante, and R.~C. Myers, {\it {Universality in fast quantum
  quenches}},  {\em JHEP} {\bf 02} (2015) 167,
  [\href{http://arxiv.org/abs/1411.7710}{{\tt arXiv:1411.7710}}].

\bibitem{Buchel:2014gta}
A.~Buchel, R.~C. Myers, and A.~van Niekerk, {\it {Nonlocal probes of
  thermalization in holographic quenches with spectral methods}},  {\em JHEP}
  {\bf 02} (2015) 017, [\href{http://arxiv.org/abs/1410.6201}{{\tt
  arXiv:1410.6201}}]. [Erratum: JHEP07,137(2015)].

\bibitem{Liu:2013iza}
H.~Liu and S.~J. Suh, {\it {Entanglement Tsunami: Universal Scaling in
  Holographic Thermalization}},  {\em Phys. Rev. Lett.} {\bf 112} (2014)
  011601, [\href{http://arxiv.org/abs/1305.7244}{{\tt arXiv:1305.7244}}].

\bibitem{Gao:2012aw}
X.~Gao, A.~M. Garcia-Garcia, H.~B. Zeng, and H.-Q. Zhang, {\it {Normal modes
  and time evolution of a holographic superconductor after a quantum quench}},
  {\em JHEP} {\bf 06} (2014) 019, [\href{http://arxiv.org/abs/1212.1049}{{\tt
  arXiv:1212.1049}}].

\bibitem{Garcia-Garcia:2013rha}
A.~M. García-García, H.~B. Zeng, and H.~Q. Zhang, {\it {A thermal quench
  induces spatial inhomogeneities in a holographic superconductor}},  {\em
  JHEP} {\bf 07} (2014) 096, [\href{http://arxiv.org/abs/1308.5398}{{\tt
  arXiv:1308.5398}}].

\bibitem{Bai:2014tla}
X.~Bai, B.-H. Lee, L.~Li, J.-R. Sun, and H.-Q. Zhang, {\it {Time Evolution of
  Entanglement Entropy in Quenched Holographic Superconductors}},  {\em JHEP}
  {\bf 04} (2015) 066, [\href{http://arxiv.org/abs/1412.5500}{{\tt
  arXiv:1412.5500}}].

\bibitem{Danielsson:1999fa}
U.~H. Danielsson, E.~Keski-Vakkuri, and M.~Kruczenski, {\it {Black hole
  formation in AdS and thermalization on the boundary}},  {\em JHEP} {\bf 02}
  (2000) 039, [\href{http://arxiv.org/abs/hep-th/9912209}{{\tt
  hep-th/9912209}}].

\bibitem{Bhattacharyya:2009uu}
S.~Bhattacharyya and S.~Minwalla, {\it {Weak Field Black Hole Formation in
  Asymptotically AdS Spacetimes}},  {\em JHEP} {\bf 09} (2009) 034,
  [\href{http://arxiv.org/abs/0904.0464}{{\tt arXiv:0904.0464}}].

\bibitem{Caceres:2014pda}
E.~Caceres, A.~Kundu, J.~F. Pedraza, and D.-L. Yang, {\it {Weak Field Collapse
  in AdS: Introducing a Charge Density}},  {\em JHEP} {\bf 06} (2015) 111,
  [\href{http://arxiv.org/abs/1411.1744}{{\tt arXiv:1411.1744}}].

\bibitem{AbajoArrastia:2010yt}
J.~Abajo-Arrastia, J.~Aparicio, and E.~Lopez, {\it {Holographic Evolution of
  Entanglement Entropy}},  {\em JHEP} {\bf 11} (2010) 149,
  [\href{http://arxiv.org/abs/1006.4090}{{\tt arXiv:1006.4090}}].

\bibitem{Ebrahim:2010ra}
H.~Ebrahim and M.~Headrick, {\it {Instantaneous Thermalization in Holographic
  Plasmas}},  \href{http://arxiv.org/abs/1010.5443}{{\tt arXiv:1010.5443}}.

\bibitem{Chesler:2009cy}
P.~M. Chesler and L.~G. Yaffe, {\it {Boost invariant flow, black hole
  formation, and far-from-equilibrium dynamics in N = 4 supersymmetric
  Yang-Mills theory}},  {\em Phys. Rev.} {\bf D82} (2010) 026006,
  [\href{http://arxiv.org/abs/0906.4426}{{\tt arXiv:0906.4426}}].

\bibitem{Garfinkle:2011tc}
D.~Garfinkle, L.~A. Pando~Zayas, and D.~Reichmann, {\it {On Field Theory
  Thermalization from Gravitational Collapse}},  {\em JHEP} {\bf 02} (2012)
  119, [\href{http://arxiv.org/abs/1110.5823}{{\tt arXiv:1110.5823}}].

\bibitem{Balasubramanian:2010ce}
V.~Balasubramanian, A.~Bernamonti, J.~de~Boer, N.~Copland, B.~Craps,
  E.~Keski-Vakkuri, B.~Muller, A.~Schafer, M.~Shigemori, and W.~Staessens, {\it
  {Thermalization of Strongly Coupled Field Theories}},  {\em Phys. Rev. Lett.}
  {\bf 106} (2011) 191601, [\href{http://arxiv.org/abs/1012.4753}{{\tt
  arXiv:1012.4753}}].

\bibitem{Balasubramanian:2011ur}
V.~Balasubramanian, A.~Bernamonti, J.~de~Boer, N.~Copland, B.~Craps,
  E.~Keski-Vakkuri, B.~Muller, A.~Schafer, M.~Shigemori, and W.~Staessens, {\it
  {Holographic Thermalization}},  {\em Phys. Rev.} {\bf D84} (2011) 026010,
  [\href{http://arxiv.org/abs/1103.2683}{{\tt arXiv:1103.2683}}].

\bibitem{Galante:2012pv}
D.~Galante and M.~Schvellinger, {\it {Thermalization with a chemical potential
  from AdS spaces}},  {\em JHEP} {\bf 07} (2012) 096,
  [\href{http://arxiv.org/abs/1205.1548}{{\tt arXiv:1205.1548}}].

\bibitem{Caceres:2012em}
E.~Caceres and A.~Kundu, {\it {Holographic Thermalization with Chemical
  Potential}},  {\em JHEP} {\bf 09} (2012) 055,
  [\href{http://arxiv.org/abs/1205.2354}{{\tt arXiv:1205.2354}}].

\bibitem{Camilo:2014npa}
G.~Camilo, B.~Cuadros-Melgar, and E.~Abdalla, {\it {Holographic thermalization
  with a chemical potential from Born-Infeld electrodynamics}},  {\em JHEP}
  {\bf 02} (2015) 103, [\href{http://arxiv.org/abs/1412.3878}{{\tt
  arXiv:1412.3878}}].

\bibitem{Giordano:2014kya}
A.~Giordano, N.~E. Grandi, and G.~A. Silva, {\it {Holographic thermalization of
  charged operators}},  {\em JHEP} {\bf 05} (2015) 016,
  [\href{http://arxiv.org/abs/1412.7953}{{\tt arXiv:1412.7953}}].

\bibitem{Zeng:2013mca}
X.~Zeng and W.~Liu, {\it {Holographic thermalization in Gauss-Bonnet gravity}},
   {\em Phys. Lett.} {\bf B726} (2013) 481--487,
  [\href{http://arxiv.org/abs/1305.4841}{{\tt arXiv:1305.4841}}].

\bibitem{Zeng:2013fsa}
X.-X. Zeng, X.-M. Liu, and W.-B. Liu, {\it {Holographic thermalization with a
  chemical potential in Gauss-Bonnet gravity}},  {\em JHEP} {\bf 03} (2014)
  031, [\href{http://arxiv.org/abs/1311.0718}{{\tt arXiv:1311.0718}}].

\bibitem{Zhang:2015dia}
S.-J. Zhang and E.~Abdalla, {\it {Holographic Thermalization in Charged Dilaton
  Anti-de Sitter Spacetime}},  {\em Nucl. Phys.} {\bf B896} (2015) 569--586,
  [\href{http://arxiv.org/abs/1503.0770}{{\tt arXiv:1503.0770}}].

\bibitem{Zhang:2014cga}
S.-J. Zhang, B.~Wang, E.~Abdalla, and E.~Papantonopoulos, {\it {Holographic
  thermalization in Gauss-Bonnet gravity with de Sitter boundary}},  {\em Phys.
  Rev.} {\bf D91} (2015), no.~10 106010,
  [\href{http://arxiv.org/abs/1412.7073}{{\tt arXiv:1412.7073}}].

\bibitem{Fonda:2014ula}
P.~Fonda, L.~Franti, V.~Keränen, E.~Keski-Vakkuri, L.~Thorlacius, and
  E.~Tonni, {\it {Holographic thermalization with Lifshitz scaling and
  hyperscaling violation}},  {\em JHEP} {\bf 08} (2014) 051,
  [\href{http://arxiv.org/abs/1401.6088}{{\tt arXiv:1401.6088}}].

\bibitem{Alishahiha:2014cwa}
M.~Alishahiha, A.~F. Astaneh, and M.~R.~M. Mozaffar, {\it {Thermalization in
  backgrounds with hyperscaling violating factor}},  {\em Phys. Rev.} {\bf D90}
  (2014), no.~4 046004, [\href{http://arxiv.org/abs/1401.2807}{{\tt
  arXiv:1401.2807}}].

\bibitem{Zeng:2014xpa}
X.-X. Zeng, X.-M. Liu, and W.-B. Liu, {\it {Holographic thermalization in
  noncommutative geometry}},  {\em Phys. Lett.} {\bf B744} (2015) 48--54,
  [\href{http://arxiv.org/abs/1407.5262}{{\tt arXiv:1407.5262}}].

\bibitem{Aref'eva:2013wma}
I.~Aref'eva, A.~Bagrov, and A.~S. Koshelev, {\it {Holographic Thermalization
  from Kerr-AdS}},  {\em JHEP} {\bf 07} (2013) 170,
  [\href{http://arxiv.org/abs/1305.3267}{{\tt arXiv:1305.3267}}].

\bibitem{Hubeny:2013dea}
V.~E. Hubeny and H.~Maxfield, {\it {Holographic probes of collapsing black
  holes}},  {\em JHEP} {\bf 03} (2014) 097,
  [\href{http://arxiv.org/abs/1312.6887}{{\tt arXiv:1312.6887}}].

\bibitem{Hartnoll:2009sz}
S.~A. Hartnoll, {\it {Lectures on holographic methods for condensed matter
  physics}},  {\em Class. Quant. Grav.} {\bf 26} (2009) 224002,
  [\href{http://arxiv.org/abs/0903.3246}{{\tt arXiv:0903.3246}}].

\bibitem{Kachru:2008yh}
S.~Kachru, X.~Liu, and M.~Mulligan, {\it {Gravity duals of Lifshitz-like fixed
  points}},  {\em Phys. Rev.} {\bf D78} (2008) 106005,
  [\href{http://arxiv.org/abs/0808.1725}{{\tt arXiv:0808.1725}}].

\bibitem{Taylor:2008tg}
M.~Taylor, {\it {Non-relativistic holography}},
  \href{http://arxiv.org/abs/0812.0530}{{\tt arXiv:0812.0530}}.

\bibitem{Chemissany:2014xsa}
W.~Chemissany and I.~Papadimitriou, {\it {Lifshitz holography: The whole
  shebang}},  {\em JHEP} {\bf 01} (2015) 052,
  [\href{http://arxiv.org/abs/1408.0795}{{\tt arXiv:1408.0795}}].

\bibitem{Gath:2012pg}
J.~Gath, J.~Hartong, R.~Monteiro, and N.~A. Obers, {\it {Holographic Models for
  Theories with Hyperscaling Violation}},  {\em JHEP} {\bf 04} (2013) 159,
  [\href{http://arxiv.org/abs/1212.3263}{{\tt arXiv:1212.3263}}].

\bibitem{Gouteraux:2012yr}
B.~Gouteraux and E.~Kiritsis, {\it {Quantum critical lines in holographic
  phases with (un)broken symmetry}},  {\em JHEP} {\bf 04} (2013) 053,
  [\href{http://arxiv.org/abs/1212.2625}{{\tt arXiv:1212.2625}}].

\bibitem{Griffin:2012qx}
T.~Griffin, P.~Hořava, and C.~M. Melby-Thompson, {\it {Lifshitz Gravity for
  Lifshitz Holography}},  {\em Phys. Rev. Lett.} {\bf 110} (2013), no.~8
  081602, [\href{http://arxiv.org/abs/1211.4872}{{\tt arXiv:1211.4872}}].

\bibitem{Donos:2010tu}
A.~Donos and J.~P. Gauntlett, {\it {Lifshitz Solutions of D=10 and D=11
  supergravity}},  {\em JHEP} {\bf 12} (2010) 002,
  [\href{http://arxiv.org/abs/1008.2062}{{\tt arXiv:1008.2062}}].

\bibitem{Chemissany:2011mb}
W.~Chemissany and J.~Hartong, {\it {From D3-Branes to Lifshitz Space-Times}},
  {\em Class. Quant. Grav.} {\bf 28} (2011) 195011,
  [\href{http://arxiv.org/abs/1105.0612}{{\tt arXiv:1105.0612}}].

\bibitem{Christensen:2013rfa}
M.~H. Christensen, J.~Hartong, N.~A. Obers, and B.~Rollier, {\it {Boundary
  Stress-Energy Tensor and Newton-Cartan Geometry in Lifshitz Holography}},
  {\em JHEP} {\bf 01} (2014) 057, [\href{http://arxiv.org/abs/1311.6471}{{\tt
  arXiv:1311.6471}}].

\bibitem{Korovin:2013bua}
Y.~Korovin, K.~Skenderis, and M.~Taylor, {\it {Lifshitz as a deformation of
  Anti-de Sitter}},  {\em JHEP} {\bf 08} (2013) 026,
  [\href{http://arxiv.org/abs/1304.7776}{{\tt arXiv:1304.7776}}].

\bibitem{Korovin:2013nha}
Y.~Korovin, K.~Skenderis, and M.~Taylor, {\it {Lifshitz from AdS at finite
  temperature and top down models}},  {\em JHEP} {\bf 11} (2013) 127,
  [\href{http://arxiv.org/abs/1306.3344}{{\tt arXiv:1306.3344}}].

\bibitem{Guica:2010sw}
M.~Guica, K.~Skenderis, M.~Taylor, and B.~C. van Rees, {\it {Holography for
  Schrodinger backgrounds}},  {\em JHEP} {\bf 02} (2011) 056,
  [\href{http://arxiv.org/abs/1008.1991}{{\tt arXiv:1008.1991}}].

\bibitem{PhysRevB.26.154}
D.~Boyanovsky and J.~L. Cardy, {\it {Critical behavior of $m$-component magnets
  with correlated impurities}},  {\em Phys. Rev. B} {\bf 26} (1982) 154--170.

\bibitem{PhysRevB.50.7526}
A.~W.~W. Ludwig, M.~P.~A. Fisher, R.~Shankar, and G.~Grinstein, {\it {Integer
  quantum Hall transition: An alternative approach and exact results}},  {\em
  Phys. Rev. B} {\bf 50} (1994) 7526--7552.

\bibitem{1998PhRvL..80.5409Y}
J.~Ye and S.~Sachdev, {\it {Coulomb Interactions at Quantum Hall Critical
  Points of Systems in a Periodic Potential}},  {\em Physical Review Letters}
  {\bf 80} (1998) 5409--5412,
  [\href{http://arxiv.org/abs/cond-mat/9712161}{{\tt cond-mat/9712161}}].

\bibitem{Herbut:2006cs}
I.~F. Herbut, {\it {Interactions and phase transitions on graphene's honeycomb
  lattice}},  {\em Phys. Rev. Lett.} {\bf 97} (2006) 146401,
  [\href{http://arxiv.org/abs/cond-mat/0606195}{{\tt cond-mat/0606195}}].

\bibitem{Son:2007ja}
D.~T. Son, {\it {Quantum critical point in graphene approached in the limit of
  infinitely strong Coulomb interaction}},
  \href{http://arxiv.org/abs/cond-mat/0701501}{{\tt cond-mat/0701501}}.

\bibitem{2000PhRvB..6114723H}
I.~F. Herbut, {\it {Critical exponents at the superconductor-insulator
  transition in dirty-boson systems}},  {\em Phys. Rev. B} {\bf 61} (2000)
  14723, [\href{http://arxiv.org/abs/cond-mat/0001040}{{\tt
  cond-mat/0001040}}].

\bibitem{Maldacena:2008wh}
J.~Maldacena, D.~Martelli, and Y.~Tachikawa, {\it {Comments on string theory
  backgrounds with non-relativistic conformal symmetry}},  {\em JHEP} {\bf 10}
  (2008) 072, [\href{http://arxiv.org/abs/0807.1100}{{\tt arXiv:0807.1100}}].

\bibitem{Deger:1998nm}
S.~Deger, A.~Kaya, E.~Sezgin, and P.~Sundell, {\it {Spectrum of D = 6, N=4b
  supergravity on AdS in three-dimensions x S**3}},  {\em Nucl. Phys.} {\bf
  B536} (1998) 110--140, [\href{http://arxiv.org/abs/hep-th/9804166}{{\tt
  hep-th/9804166}}].

\bibitem{Biran:1983iy}
B.~Biran, A.~Casher, F.~Englert, M.~Rooman, and P.~Spindel, {\it {The
  Fluctuating Seven Sphere in Eleven-dimensional Supergravity}},  {\em Phys.
  Lett.} {\bf B134} (1984) 179.

\bibitem{Kim:1985ez}
H.~J. Kim, L.~J. Romans, and P.~van Nieuwenhuizen, {\it {The Mass Spectrum of
  Chiral N=2 D=10 Supergravity on S**5}},  {\em Phys. Rev.} {\bf D32} (1985)
  389.

\bibitem{deHaro:2000vlm}
S.~de~Haro, S.~N. Solodukhin, and K.~Skenderis, {\it {Holographic
  reconstruction of space-time and renormalization in the AdS / CFT
  correspondence}},  {\em Commun. Math. Phys.} {\bf 217} (2001) 595--622,
  [\href{http://arxiv.org/abs/hep-th/0002230}{{\tt hep-th/0002230}}].

\bibitem{Ross:2009ar}
S.~F. Ross and O.~Saremi, {\it {Holographic stress tensor for non-relativistic
  theories}},  {\em JHEP} {\bf 09} (2009) 009,
  [\href{http://arxiv.org/abs/0907.1846}{{\tt arXiv:0907.1846}}].

\bibitem{Bianchi:2001kw}
M.~Bianchi, D.~Z. Freedman, and K.~Skenderis, {\it {Holographic
  renormalization}},  {\em Nucl. Phys.} {\bf B631} (2002) 159--194,
  [\href{http://arxiv.org/abs/hep-th/0112119}{{\tt hep-th/0112119}}].

\bibitem{Booth:2005qc}
I.~Booth, {\it {Black hole boundaries}},  {\em Can. J. Phys.} {\bf 83} (2005)
  1073--1099, [\href{http://arxiv.org/abs/gr-qc/0508107}{{\tt gr-qc/0508107}}].

\bibitem{Ryu:2006bv}
S.~Ryu and T.~Takayanagi, {\it {Holographic derivation of entanglement entropy
  from AdS/CFT}},  {\em Phys. Rev. Lett.} {\bf 96} (2006) 181602,
  [\href{http://arxiv.org/abs/hep-th/0603001}{{\tt hep-th/0603001}}].

\bibitem{Hubeny:2007xt}
V.~E. Hubeny, M.~Rangamani, and T.~Takayanagi, {\it {A Covariant holographic
  entanglement entropy proposal}},  {\em JHEP} {\bf 07} (2007) 062,
  [\href{http://arxiv.org/abs/0705.0016}{{\tt arXiv:0705.0016}}].

\bibitem{Liu:2012eea}
H.~Liu and M.~Mezei, {\it {A Refinement of entanglement entropy and the number
  of degrees of freedom}},  {\em JHEP} {\bf 04} (2013) 162,
  [\href{http://arxiv.org/abs/1202.2070}{{\tt arXiv:1202.2070}}].

\end{thebibliography}\endgroup
\end{document}